\documentclass[12pt]{article}
\pdfoutput=1
\usepackage[latin1]{inputenc}
\usepackage{amsmath}
\usepackage{amsfonts}
\usepackage{amssymb}
\usepackage{graphicx,psfrag}
\usepackage{geometry}
\usepackage{amssymb,epsfig,subfigure}
\usepackage{hyperref}
\usepackage{enumerate}

\def\pplogo{\vbox{\kern-\headheight\kern -29pt
\halign{##&##\hfil\cr&{\ppnumber}\cr\rule{0pt}{2.5ex}&\ppdate\cr}}}
\makeatletter
\def\ps@firstpage{\ps@empty \def\@oddhead{\hss\pplogo}%
  \let\@evenhead\@oddhead 
}
\def\maketitle{\par
 \begingroup
 \def\thefootnote{\fnsymbol{footnote}}
 \def\@makefnmark{\hbox{$^{\@thefnmark}$\hss}}
 \if@twocolumn
 \twocolumn[\@maketitle]
 \else \newpage
 \global\@topnum\z@ \@maketitle \fi\thispagestyle{firstpage}\@thanks
 \endgroup
 \setcounter{footnote}{0}
 \let\maketitle\relax
 \let\@maketitle\relax
 \gdef\@thanks{}\gdef\@author{}\gdef\@title{}\let\thanks\relax}
\makeatother

\numberwithin{equation}{section}

\newcommand{\mathd}{\mathrm{d}}
\newcommand{\mathD}{\mathrm{D}}

\newcommand{\tmop}[1]{\ensuremath{\operatorname{#1}}}

\renewcommand{\dag}{\dagger}

\newcommand{\be}{\begin{equation}}
\newcommand{\bea}{\begin{eqnarray}}
\newcommand{\ee}{\end{equation}}
\newcommand{\eea}{\end{eqnarray}}

\newcommand{\mc}{\mathcal}

\renewcommand{\t}{\tilde}

\newcommand{\Biglp}{\Bigl(}
\newcommand{\Bigrp}{\Bigr)}
\newcommand{\TeXmacs}{T\kern-.1667em\lower.5ex\hbox{E}\kern-.125emX\kern-.1em\lower.5ex\hbox{\textsc{m\kern-.05ema\kern-.125emc\kern-.05ems}}}

\begin{document}

\setcounter{page}0
\def\ppnumber{\vbox{\baselineskip14pt
}}
\def\ppdate{\footnotesize{SLAC-PUB-14284}} \date{}

\author{\normalsize Ben Heidenreich,$^{1}$ Liam McAllister,$^{1}$ and Gonzalo Torroba$^{2}$\\
[7mm]
{\normalsize $^1$ \it Department of Physics, Cornell University, Ithaca, NY 14853 USA}\\
{\normalsize $^2$ \it  SLAC and Department of Physics, Stanford University, Stanford, CA 94309 USA}\\
}

\bigskip

\title{\bf \Large Dynamic SU(2) Structure from Seven-branes}
\vskip 14pt
\maketitle





\noindent We obtain a family of supersymmetric solutions of type IIB supergravity with dynamic $SU(2)$ structure, which describe
the local geometry near a stack of four D7-branes and one O7-plane wrapping a rigid four-cycle.  The deformation to a generalized complex geometry is interpreted as a consequence of nonperturbative effects in the seven-brane gauge theory. We formulate the problem for seven-branes wrapping the base of an appropriate del Pezzo cone, and in the near-stack limit in which the four-cycle is flat, we obtain an exact solution in closed form.
Our solutions serve to characterize the local geometry of nonperturbatively-stabilized flux compactifications.

\vfil
\begin{flushleft}
\small \today
\end{flushleft}

\bigskip
\newpage

\tableofcontents

\vskip 1cm

\section{Introduction}\label{sec:intro}

Two fundamental goals in string theory are characterizing the vacua of the theory
and understanding strongly-coupled four-dimensional gauge theories from a ten-dimensional viewpoint.  These problems intersect when the dynamics of
a strongly-coupled four-dimensional gauge theory determines
the potential for compactification moduli, as in flux compactifications of type IIB string theory, where gaugino condensation on seven-branes provides an important contribution to the potential for the K\"ahler moduli.

In this work we present a family of explicit local solutions that describe the region near a stack of seven-branes wrapping a rigid four-cycle.  We argue that a subclass of our solutions encode seven-brane nonperturbative effects in ten-dimensional supergravity. We begin in \S\ref{subsec:introcompact} by motivating the study of seven-brane gaugino condensation, then explain in \S\ref{subsec:introGCG} why the corresponding solution will be a generalized complex geometry.

\subsection{Gaugino condensation in string compactifications}\label{subsec:introcompact}

In a compactification of type IIB string theory on a Calabi-Yau
threefold, classical vacua involving nonvanishing fluxes and localized
D-brane and orientifold plane sources provide a rich array of
four-dimensional theories with ${\cal N} =1$ or ${\cal N} =0$
supersymmetry.
The K\"ahler moduli are typically unfixed in the classical vacuum and mediate gravitational-strength
interactions that preclude the construction of realistic models of particle physics and cosmology.  Perturbative and nonperturbative effects may be
expected to give mass to the K\"ahler moduli, and in certain special
cases the dominant effects can be computed.

The proposals for K\"ahler moduli stabilization of \cite{Kachru:2003aw, BBCQ, Braun}
incorporate nonperturbative effects arising from branes wrapping
four-cycles in the compact space \cite{Witten}: these can be either Euclidean D3-branes, or
a stack of $(p,q)$ seven-branes giving
rise to a four-dimensional gauge theory that is strongly coupled in
the infrared and generates a nonperturbative superpotential.
Considerable efforts have
been directed at understanding the four-dimensional
effective theory incorporating the classical flux superpotential and
the nonperturbative superpotential arising from wrapped seven-branes,
but the corresponding ten-dimensional configuration that encodes the
effects of nonperturbative dynamics on the seven-branes remains mysterious.

Why understand the imprint of four-dimensional nonperturbative physics
in ten dimensions, given that the four-dimensional theory
itself is well-understood?  One powerful motivation comes from the utility of higher-dimensional locality in model building.
For example, one can construct a supersymmetric visible sector
on D-branes in one region of a compactification, and later incorporate
soft terms induced by supersymmetry breaking in a distant region.  A
prerequisite for such an analysis is locality in the compact space,
which manifestly requires a ten-dimensional solution.   Thus,
understanding locality in nonperturbatively-stabilized vacua requires a ten-dimensional solution
encoding the effects that stabilize the K\"ahler
moduli.  A second motivation is that local geometries describing strong gauge dynamics on seven-branes could be glued into compact geometries as `modules' effecting the stabilization of K\"ahler moduli.
A third important motivation is that gravity solutions can shed new light on the dynamics of supersymmetric gauge theories on seven-branes via gauge-gravity duality, as we explain in \S\ref{subsec:introGCG}.

We are therefore led to search for explicit local solutions describing
strong gauge dynamics on seven-branes wrapping rigid four-cycles.
A natural class of local Calabi-Yau geometries containing rigid four-cycles are complex cones
over del Pezzo surfaces.  The possibilities for wrapping seven-branes on the del Pezzo base of
such a cone are quite constrained: the total seven-brane charge must
vanish (see \S\ref{subsec:orient}). A convenient choice obeying this constraint is a stack of four
D7-branes that coincide with an O7-plane: the total seven-brane charge
and tension vanish, but lower-dimensional brane charges may be induced
on the stack, e.g.\ by $\alpha^{\prime}$ corrections.\footnote{In fact, negative D3-brane charge and tension induced on the seven-branes
will contribute to the formation of a singularity near the four-cycle.}
The resulting four-dimensional gauge theory is pure\footnote{See \S\ref{subsec:various} for important comments on how the global anomaly of~\cite{Minasian:1997mm,Freed:1999vc} may constrain the resulting gauge group.  For simplicity of presentation we will speak of gaugino condensation in pure super Yang-Mills throughout, while keeping in mind that the full gauge theory may be more complicated.} super Yang-Mills, which exhibits gaugino condensation and chiral symmetry breaking at low energies.

We conclude that a promising setting for studying the backreaction of
four-dimensional nonperturbative effects is a Calabi-Yau cone whose
base, a del Pezzo surface, is the fixed-point locus of an orientifold
action and in addition is wrapped by four D7-branes.  In this work we
formulate the problem in this general setting but provide a detailed
solution for a simpler special case, in which one zooms in on the
region near the seven-branes.  Results for del Pezzo cones will be
provided elsewhere.

\subsection{Nonperturbative effects from seven-branes and generalized complex geometry}\label{subsec:introGCG}

The above configuration of seven-branes will preserve
four supercharges that are embedded in the type IIB ten-dimensional Majorana-Weyl supersymmetry generators $\epsilon^i$ as
\be\label{eq:10dspinor}
\epsilon^i = \zeta_+ \otimes \eta_+^i + \zeta_- \otimes \eta_-^i\;\;,\;\;i=1,2\,,
\ee
where the conventions are those of~\cite{Grana:2005sn}: $\zeta_+$ is a positive-chirality four-dimensional spinor that generates four-dimensional $\mathcal{N}=1$ supersymmetry transformations, and the $\eta_+^i$ are fixed positive-chirality six-dimensional spinors, with $\eta_-^i, \zeta_-$ the Majorana conjugates of $\eta_+^i, \zeta_+$, respectively.

Taken as strictly classical sources, four D7-branes wrapping the del Pezzo fixed-point locus of an orientifold will preserve `type B' supersymmetry \cite{Grana:2005sn}, with $\eta_+^1 = \pm i \eta_+^2$.  However, a supergravity solution encoding gaugino condensation requires a different supersymmetry, as we now argue.  In a type B background (as characterized e.g.\ in~\cite{Giddings:2001yu}) a D3-brane experiences vanishing potential, while gaugino condensation on seven-branes is known to lift the D3-brane moduli space.   Further evidence comes from~\cite{BDKKM}, where it was established that in a type B solution, gaugino condensation from D7-branes on a rigid cycle sources imaginary anti-self-dual flux, of Hodge type (1,2), which is incompatible with the supersymmetry of the background.  In fact, a supergravity solution encoding seven-brane gaugino condensation requires that the  internal spinors $\eta_+^1$ and $\eta_+^2$  be distinct, as argued in \S\ref{sec:SUSY}, and a direct analysis of the dilatino and gravitino variations becomes more involved. Mathematically, the two internal spinors define a \textit{local} or \textit{dynamic} $SU(2)$ structure that can be dealt with most efficiently in terms of generalized complex geometry~\cite{Grana:2005sn}, as originally proposed for D7-brane gaugino condensation in \cite{Koerber:2007xk}.\footnote{An earlier example that displays gaugino condensation and $(1,2)$ three-form flux is the Polchinski-Strassler solution~\cite{Polchinski:2000uf}; the ten-dimensional analysis of~\cite{Grana:2000jj,Lopes Cardoso:2004ni} revealed the presence of an $SU(2)$ structure.}

Our goal in this work is to investigate supergravity solutions with dynamic $SU(2)$ structure that describe the gauge theory dynamics of compact seven-branes.
The approach taken here is purely ten-dimensional, and we do not assume the existence of a four-dimensional nonperturbative superpotential.
Rather, in the spirit of~\cite{Polchinski:2000uf,Klebanov:2000hb}, the supergravity solution describing seven-branes wrapped on an appropriate rigid cycle should already encode the effects of gaugino condensation.\footnote{See \cite{Andrew} for a ten-dimensional description of gaugino condensation in the heterotic string.}

Let us comment on the relation between our approach and well-understood AdS/CFT descriptions of gaugino condensation in other systems.  The gravity dual of pure super Yang-Mills is not known, and is expected to correspond to a regime of large curvature and strong corrections to classical supergravity.  However, pure super Yang-Mills can be embedded into a branch of a large N quiver theory, or into a higher-dimensional gauge theory, leading to well-defined supergravity solutions, albeit with extra fields that are absent from the pure glue theory.

A celebrated example is the Klebanov-Strassler solution~\cite{Klebanov:2000hb}, in which an $SU(N) \times SU(N+M)$ quiver theory, which confines and breaks chiral symmetry in the infrared,
is dual to a conformally-Calabi-Yau solution, the warped deformed conifold. In this case the boundary supersymmetry is of type B, facilitating embedding in well-understood flux compactifications \cite{Giddings:2001yu}.
In contrast, there is no known supergravity solution dual to a gauge theory with seven-branes wrapped on a rigid four-cycle.   One obstacle is that taking a four-dimensional limit by shrinking the four-cycle generically gives chiral matter and an anomalous gauge theory.\footnote{Anomaly-free gauge theories from seven-branes on del Pezzo cones were constructed recently in~\cite{Franco:2010jv}.} The anomaly has to be canceled by adding extra ingredients, such as orientifolds, but
these do not decouple from the low energy theory and complicate both the gauge theory analysis and the supergravity solution.
Furthermore, the number of seven-branes cannot be taken to be large, so that it is difficult
to obtain parametric control of the curvatures appearing on the gravity side.

One might expect that our noncompact supergravity solutions should
capture the backreaction associated to the gauge theory dynamics. However, there is  at present no fully-realized gauge-gravity duality for our system: our solutions do not include a large number of D3-branes, and the asymptotic geometry is very different from $AdS_5$.
We will suggest that our solutions
describe the behavior of the gauge theory in the deep infrared, after most of the degrees of freedom have renormalized away. Introducing a large number of color branes might yield a solution in which a more precise gauge/gravity dictionary can be constructed, but we leave this question for the future.

\subsection{Overview}\label{subsec:overview}

Our main result is an explicit supersymmetric solution with dynamic (and in general, type-changing) $SU(2)$ structure, which
describes the generalized complex geometry near a stack of four D7-branes and one O7-plane.  This solution arises as a limiting case of configurations in which the seven-branes wrap a compact four-cycle: in the example we provide,
this cycle is the base of the Calabi-Yau cone over $\mathbb{P}^2$, ${\cal O}(-3)_{\mathbb P^2}$.
We construct our solution in closed form after solving the supersymmetry conditions for an
$AdS_{4}$ vacuum in the limit of vanishing cosmological constant.
The form of the supersymmetry conditions used here is $SL(2,\mathbb{R})$ covariant, which is very useful in classifying the resulting solutions.

The organization of this paper is as follows.  In \S\ref{sec:background}, we present some essential geometric background for our analysis.
In \S\ref{subsec:flat}, we describe a simple ansatz that will be our primary focus. In \S\ref{subsec:P2}, we extend our considerations to the $\mathbb{P}^2$ cone, and show that the ansatz of \S\ref{subsec:flat} arises in a scaling limit that zooms in on the $\mathbb{P}^2$.  In \S\ref{sec:SUSY}, we briefly review dynamic $SU(2)$ structure, then
present the full supersymmetry conditions for compactification to $AdS_4$ in an $SL(2,\mathbb{R})$ covariant form.  In \S\ref{sec:solution} we solve these conditions, as well as the equations of motion, to obtain the most general `AdS-like' supersymmetric solution to our ansatz. In \S\ref{sec:solprops} we describe the regime of validity of supergravity and discuss a few key geometric properties of our solutions.
In \S\ref{sec:future} we present a preliminary analysis of the relation between the above solutions and nonperturbative effects on seven-branes, and
indicate a few interesting applications and directions for future research. We conclude in \S\ref{sec:concl}. The equations of motion for our ansatz are assembled in Appendix~\ref{appendix:eoms}.

\section{Seven-branes and Orientifolds}
\label{sec:background}

We will study a stack of four D7-branes atop an O7-plane, wrapping a rigid four-cycle in a local Calabi-Yau geometry and preserving four supercharges.
The specific
local geometries that we will consider are resolved Calabi-Yau cones over del Pezzo surfaces.
In each case, the del Pezzo base is a rigid shrinking divisor within the Calabi-Yau. These properties make Calabi-Yau cones over del Pezzos the natural choice for the purpose of obtaining local gauge theories that undergo gaugino condensation. In this work we will focus primarily on the simplest del Pezzo surface, $\mathbb P^2$, but we begin with topological considerations that are valid for any del Pezzo cone, and then proceed to describe the geometry of the resolved cone over $\mathbb P^2$, ${\cal O}(-3)_{\mathbb P^2}$.

\subsection{Orientifolds of resolved del Pezzo cones}
\label{subsec:orient}

Wrapping seven-branes over the del Pezzo surface yields a supersymmetric field theory in four dimensions.   In order to obtain a pure glue theory in which gaugino condensation can occur, we do not add seven-branes wrapping non-compact divisors.  Then, owing to the rigidity of the del Pezzo within the Calabi-Yau, the four-dimensional field theory has no light matter, and develops a gaugino condensate in the infrared. One can imagine altering this situation in various ways to remove the gaugino condensate, for instance by adding D3-branes close to the seven-branes, giving light 3-7 strings. Thus, it makes sense to consider the supergravity background both with and without the effects of gaugino condensation.
We first consider the situation without a condensate, in which case the background contains only D3-branes and seven-branes, and can be studied using F-theory.

The only compact divisor in these cones is the del Pezzo itself. This places a strong constraint on the allowed brane content consistent with tadpole cancellation, as we now show. Since stacks of seven-branes carry an $SL(2,\mathbb{Z})$ monodromy depending on their total charge, we consider the $SL(2,\mathbb{Z})$ monodromy structure of the solution. For a manifold $M$ without branes, the allowed monodromies are given by homomorphisms $\Lambda : \pi_1(M) \rightarrow SL(2,\mathbb{Z})$.
Seven-branes may be thought of as topological defects in the type IIB vacuum; thus, the seven-brane charges and monodromy structures associated with a given configuration of branes are classified by homomorphisms $\Lambda : \pi_1(M') \rightarrow SL(2,\mathbb{Z})$, where $M'$ is given by $M$ minus the worldvolumes of the seven-branes.

A del Pezzo cone may be viewed as a real cone $M$
over a Sasaki-Einstein manifold which we refer to figuratively as the `horizon.' Upon excising the tip, the resulting $M'$ is homotopically equivalent to its horizon. Thus, the allowed seven-brane charges are constrained by the fundamental group of the horizon. It is known that these horizons are always simply connected up to torsion, and the only torsion groups
that appear are $\mathbb{Z}_3$ (for the $\mathbb{P}^2$ cone) and $\mathbb{Z}_2$ (for the $\mathbb{P}^1\times\mathbb{P}^1$ cone)~\cite{Liu:1998dra}. Except in these special cases, we must cancel the D7-brane tadpole \emph{locally}.\footnote{This can be seen from the field theory perspective as well: the gauge theories corresponding to disallowed configurations of seven-branes will be rendered inconsistent by anomalies~\cite{Franco:2010jv}.} A well-understood way to do this is to wrap eight D7-branes on the four-cycle and then orientifold by a $\mathbb{Z}_2$ involution whose fixed point locus is the cycle itself.

It is important to distinguish between the `upstairs' and `downstairs' geometries of the resulting orientifold. From the perspective of perturbative string theory, it is natural to do computations in the upstairs geometry, in which we have eight D7-branes coinciding with an O7-plane wrapping the base of a resolved del Pezzo cone. At energies below the Kaluza-Klein scale the resulting gauge theory is pure glue $\mathcal{N} = 1$
super Yang-Mills, where all open string fields arise from 7-7 strings and live in the adjoint representation. Classically, the composite object carries \emph{zero} seven-brane charge and tension, and generates no deficit angle.

From the perspective of F-theory, on the other hand, it is more natural to work in the downstairs geometry, where we identify under the involution to obtain a $\mathbb{Z}_2$ orbifold of the del Pezzo cone which carries a $\mathbb{Z}_2$ monodromy coming from the $-1 \in SL(2,\mathbb{Z})$. The eight D7-branes of the upstairs geometry are reduced to four by this identification. In addition, the orientifold plane appears in F-theory as a combination of \emph{two} coincident $(p,q)$ seven-branes that
separate at strong coupling~\cite{Sen:1996vd}. Thus, in the downstairs geometry, we obtain a stack of \emph{six} $(p,q)$ seven-branes. Each carries a deficit angle of $\pi/6$, for a total deficit angle of $\pi$, as required to match the deficit angle of $({\cal O}(-3)_{\mathbb P^2})/{{\mathbb Z}_2}$.

Although seven-branes are often conveniently treated in F-theory, the more general
supersymmetric backgrounds
that will be relevant in our analysis are not well-studied in F-theory,
and we opt to work in ten-dimensional supergravity. (In \S\ref{subsec:validity} we will explicitly demonstrate parametric control of the supergravity approximation.)
We will generally work in the upstairs geometry, removing modes forbidden by the involution.

\subsection{The Calabi-Yau geometry of the $\mathbb{P}^2$ cone} \label{subsec:CYgeom}

We now review the well-known geometry of the Calabi-Yau cone over $\mathbb{P}^2$, i.e.\ the resolution of the orbifold $\mathbb{C}^3 /\mathbb{Z}_3$, where
the $\mathbb{Z}_3$ acts by
\be
  z^i \rightarrow e^{2 \pi i / 3} z^i
\ee
on the $\mathbb{C}^3$ coordinates $z^i$. The Calabi-Yau metric has a $U(3)=SU(3)\times U(1)_{\psi}$ isometry
that acts naturally on the $z^i$, where the $SU(3)$ subgroup acts on the $\mathbb{P}^2$ base in the natural way, and we normalize the $U(1)_{\psi}$ such that the $z^i$ carry charge $+1/3$. Let us define the $U(3)$ invariant radial coordinate
\be\label{eq:defr}
\rho^2 \equiv \sum_{i=1}^3 |z^i|^2\,.
\ee
Each surface of constant $\rho$ is diffeomorphic to the horizon, $S^5/\mathbb Z_3$. It is useful to think of this space as a Hopf fibration over $\mathbb P^2$, where
$U(1)_{\psi}$ rotates the fibers.
Locally, we can define a circle coordinate $\psi$ with periodicity $2\pi$, such that
$e^{i \psi^{(i)}/3} = z^i/|z^i|$ for some arbitrarily chosen $i$, and $U(1)_{\psi}$ rotations are equivalent to shifts in $\psi$.

The $\mathbb{Z}_3$ orbifold singularity may be resolved into a $\mathbb{P}^2$. Viewed from the $\mathbb{C}^3/\mathbb{Z}^3 -\{0\}$ region, the size of the resolution is visible as a normalizable perturbation to the conical metric, as we now review. The conical Calabi-Yau metric for this space, either singular or resolved,  can be obtained by taking the K\"ahler potential $K$ to depend only on $\rho$, so that (cf. e.g.~\cite{Krishnan:2008kv})
\be\label{eq:Kahler}
 g_{mn}(y) dy^m dy^n = \partial_i \Bigl( z_{\bar{j}} K'(\rho^2)\Bigr)\,dz^i d \bar{z}^{\bar{j}} =  K'(\rho^2) \, \sum_i dz^i d \bar{z}^{\bar{i}}+ K''(\rho^2)\,\sum_{i,j}\bar{z}_i z_{\bar{j}} dz^i d \bar{z}^{\bar{j}}\,.
\ee
Here primes denote derivatives with respect to $\rho^2$ and $z_{\bar{i}} \equiv \delta_{\bar{i} j} z^j$. The Ricci-flatness condition ${\rm det} (\partial_i \bar{\partial}_ j K) = 1$ admits the solution
\be\label{eq:solK}
\rho^2 K'(\rho^2) = (\rho^6+\rho_0^6)^{1/3}\,.
\ee
Plugging (\ref{eq:solK}) into (\ref{eq:Kahler}) gives the metric on the resolved cone, where the origin $\rho=0$ has been blown up into a finite $\mathbb P^2$ with size controlled by $\rho_0$.

We wrap eight D7-branes on the $\mathbb{P}^2$ base and orientifold under the
involution
\be
\sigma : z^i \rightarrow - z^i\,,
\ee
combined with $-1 \in SL(2,\mathbb{Z})$. This involution is holomorphic and reverses the holomorphic three-form
\be \label{eqn:Omega3}
\Omega \propto \epsilon_{i j k}\, \mathd z^i \wedge \mathd z^j \wedge \mathd z^k\,,
\ee
as required to preserve $\mathcal{N}=1$ supersymmetry. The fixed point locus is the $\mathbb{P}^2$ itself. Since the stack carries no D7-brane charge or tension in the upstairs geometry, the Calabi-Yau metric we have just derived remains valid. However, there is a net D3-brane charge and tension induced by $\alpha'$ corrections to the D7-brane action~\cite{Bershadsky:1995qy}. For the del Pezzo orientifold
that we consider, the results of~\cite{Denef,Blumenhagen} imply\footnote{There are subtleties here
involving
cancellation of the global anomaly found in~\cite{Minasian:1997mm,Freed:1999vc}; cf. \S\ref{subsec:various}.}
that there is an induced negative
D3-brane charge proportional to the Euler characteristic of the base. There is also a corresponding negative tension.
These charges will backreact on the warp factor and $C_4$ in a manner which we derive in \S\ref{sec:solution}.

Since the $\mathbb{P}^2$ is rigid, we expect gaugino condensation to ensue at low energies. The condensate will source imaginary anti-self-dual fluxes~\cite{BDKKM}, and therefore the background will no longer be conformally Calabi-Yau. Our approach to this problem is to search for new supersymmetric supergravity solutions as candidates for the backreaction from nonperturbative effects, in some region away from the branes where the supergravity approximation is valid.  Let us reiterate that we do not directly incorporate gaugino condensation as a localized source (cf. \cite{Koerber:2007xk,BDKKM}), but instead obtain supergravity solutions that are consistent with the possibility of such backreaction.

The holomorphic three-form (\ref{eqn:Omega3}) is charged under $U(1)_{\psi}$, which will therefore be an R-symmetry in the gauge theory.  The R-charge of
$\Omega$
is the same as that of the four-dimensional superpotential, $R(\Omega)=+2$, so $e^{i \beta} \in U(1)_R$ corresponds to $e^{2 i \beta} \in U(1)_{\psi}$. As in the gauge theory, we expect that $U(1)_R$ is anomalous, breaking to a discrete subgroup.
Nonzero $G_3$ flux will further break the
R-symmetry since $G_3$ must be odd under the subgroup generated by the spatial involution, $z^i \rightarrow -z^i$. We will see in \S\ref{subsec:various} that the R-symmetry breaks to $\mathbb{Z}_2$ for our $SU(2)$ structure solutions. We anticipate that, in regions of parameter space where our solution provides a gravity dual to gaugino condensation, this will be the gravity analogue of the expected spontaneous R-symmetry breaking in the gauge theory due to the expectation value of the gaugino bilinear. We assume that the backreacted solution does not break the remaining $SU(3)$ symmetry.

\section{Ansatz for the Supergravity Solution}\label{sec:ansatz}

To simplify the problem, we will first focus on a small $\mathbb{R}^6$ patch near the stack. The supergravity ansatz for this limit is presented in \S \ref{subsec:flat}. In \S\ref{subsec:P2} we present the supergravity ansatz for the full Calabi-Yau cone over $\mathbb{P}^2$, and in \S\ref{subsec:nearstacklimit} we show that the two ans\"atze are related by a particular near-stack limit.

\subsection{The near-stack region}\label{subsec:flat}

We consider the $\mathbb{R}^6$ neighborhood of a small piece of the seven-brane
stack,
and approximate the stack as flat within this region. Imposing the $SU(3)$ symmetry group, the full geometry can be recovered from its local form in the $\rho \ll \rho_0$ limit (where $\rho_0$ determines the size of $\mathbb P^2$ as in (\ref{eq:solK})).

The first step is to define the correct `near-stack limit'.
We focus on the decomposition of the $U(3)=SU(3)\times U(1)_{\psi}$ isometry group in this limit, leaving a more detailed mapping of the supergravity fields to \S\ref{subsec:nearstacklimit}. Consider the region $z^3\ne 0$. We define coordinates
\be \label{eqn:nearstackchart}
u^a \equiv z^a/z^3 \;\;,\;\;a=1,2 \;\;,\;\; z\equiv\frac{1}{3} (z^3)^3    \,,
\ee
which are invariant under the $\mathbb{Z}_3$ orbifold action, and carry charges $0$ and $+1$, respectively, under the $U(1)_{\psi}$. The $SU(3)$ decomposes into $SU(2)\times U(1)_T$, where the $SU(2)$ acts naturally on the $u^a$, and the $U(1)_T$ takes the form
\be
u^a \rightarrow e^{i \theta} u^a \;\;,\;\; z \rightarrow e^{-2 i \theta} z \,.
\ee
In addition, there are four generators of $SU(3)$ that
mix the $u^a$
and $z$. These take the infinitesimal form
\be
z^a \rightarrow z^a + \theta^a z^3 +  {\cal{O}}(\theta^2)\;\;,\;\;z^3 \rightarrow z^3 - \bar{\theta}_{b} z^b + {\cal{O}}(\theta^2)\,,
\ee
for complex $\theta^a$, where $\theta_{\bar{b}} \equiv \delta_{\bar{b} a} \theta^a$. Thus,
\be
u^a \rightarrow u^a \left(1 + u^b  \bar{\theta}_b \right) + \theta^a + {\cal{O}}(\theta^2) \;\;,\;\; z \rightarrow z \left( 1 - 3 u^b  \bar{\theta}_b \right) + {\cal{O}}(\theta^2)\,.
\ee
This is a nonlinear transformation on the coordinates, even at first order in $\theta^a$. However, if we additionally approximate that $r_u \equiv \sum_a |u^a|^2 \ll 1$, then we obtain
\be
u^a \rightarrow u^a + \theta^a +{\cal{O}}(r_u \theta, \theta^2)\;\;,\;\; z\rightarrow z+{\cal{O}}(r_u \theta, \theta^2) \,,
\ee
corresponding to $\mathbb{C}^2$ translations on the $u^a$. To accomplish this formally, we rescale
\be \label{eqn:nearstacklimit}
u^a \rightarrow \varepsilon u^a \;\;,\;\; z \rightarrow \varepsilon z\,,
\ee
and then truncate to leading order in $\varepsilon$.

We take (\ref{eqn:nearstacklimit}) as the initial definition of the near-stack limit. (In \S\ref{subsec:validity} we obtain a more precise definition (\ref{eqn:NSprebound1}) in terms of the effective codimension of the stack: in the near-stack limit, the seven-branes are real codimension two sources, while at longer distances they appear as real codimension six sources.)  Geometrically, this limit corresponds to zooming in on a small neighborhood of a specified patch of the D7/O7 stack.  As we have shown, `small' $SU(3)$ transformations -- those which map the small neighborhood to itself -- decompose locally into $\mathbb{C}^2 \rtimes (SU(2)\times U(1)_T)$ transformations; different local patches are related by `large' $SU(3)$ transformations.

The stack of seven-branes is located in the plane $z=0$, and the involution takes $z\rightarrow -z$. We choose a circle coordinate $\psi$ locally such that $z=r _z e^{i \psi}$ where $r_z$ is real. In the near-stack limit, we find
\be
r_z \approx \frac{1}{3} \rho^3\equiv r\,,
\ee
where $r$ is an alternate radial coordinate on the $\mathbb{P}^2$ cone that will be useful below. Thus, $r$ and $r_z$ match in the near-stack limit, and except where the distinction is important, we will
denote both by $r$.

In our notation, the low energy effective action{\footnote{The equations of motion must be supplemented by the
self-duality constraint $\tilde{F}_5 = \star_{10} \tilde{F}_5$.}} for type IIB
string theory written in Einstein frame is
\begin{eqnarray} \label{eqn:SUGRAaction}
  S & = & \frac{1}{2 \kappa_{10}^2} \int \mathd^{10} x \sqrt{- g^{(10)}}  \left[ {\cal{R}} -
  \frac{1}{2}  \left( (\nabla \phi)^2 + e^{- \phi} |H_3 |^2 + e^{2
  \phi} |F_1 |^2 + e^{\phi} | \tilde{F}_3 |^2 + \frac{1}{2} |
  \tilde{F}_5 |^2 \right) \right] \nonumber \\
  &  & - \frac{1}{4 \kappa_{10}^2} \int C_4 \wedge H_3 \wedge F_3\,,
\end{eqnarray}
where $|F_p |^2 \equiv \frac{1}{p!} F^{M_1 \ldots M_p} F^{\star}_{M_1 \ldots
M_p}$, $H_3 = \mathd B_2$, $F_p = \mathd C_{p - 1}$, and
\be
  \tilde{F}_3 = F_3 - C_0 H_3 \;\;,\;\;
  \tilde{F}_5 = F_5 - \frac{1}{2} C_2 \wedge H_3 + \frac{1}{2} B_2 \wedge
  F_3\,.
\ee
We adopt a warped ansatz, with ten-dimensional metric
\be\label{eqn:10dmetric}
  \mathd s_{10}^2 = e^{2 A (y)}\, h_{\mu \nu} (x)\, \mathd x^{\mu} \mathd
  x^{\nu} + e^{- 2 A (y)}\, g_{m n} (y)\, \mathd y^m \mathd y^n\,,
\ee
and five-form field strength\footnote{With our sign conventions, the Hodge star associated with a $D$-dimensional metric $g$ with volume form $\Omega_{(g)}$ is defined by $\star \Biglp \mathd x^{m_1}\wedge\ldots\wedge\mathd x^{m_p} \Bigrp  = \frac{1}{(D-p)!}\, \Omega_{(g)}^{m_1 \ldots m_p}{}_{m_{p+1}\ldots m_D}  \Biglp \mathd x^{m_{p+1}}\wedge\ldots\wedge\mathd x^{m_D} \Bigrp $.}
\be\label{eqn:F5ansatz}
  \tilde{F}_5 = \left( 1 + \star_{10} \right) \Omega_4 \wedge \mathd
  \left[\alpha(y)\right]\,,
\ee
where $\alpha$ is a scalar related to $C_4$, $h_{\mu \nu} (x)$ is a maximally symmetric metric on $\mathbb{R}^{3,1}$ with cosmological constant $\Lambda={\cal{R}}_{(4)}/4$ and volume form $\Omega_4$, and $g_{m n}(y)$ (times the conformal factor $e^{-2A}$) gives the internal space metric.  We do not assume that $g_{m n}(y)$ is Calabi-Yau, or indeed even complex.  The axiodilaton $\tau(y) = C_0 + i e^{- \phi} = \tau_1 + i\, \tau_2$ varies over the compact space, and the three-form flux
\be
G_3 \equiv \frac{1}{\sqrt{\tau_2}}\, (F_3-\tau H_3)
\ee
points along the internal directions only.

Backreaction from the seven-branes changes the metric and sources various supergravity fields. We consider the most general ansatz compatible with the assumed symmetry group $\mathbb{C}^2 \rtimes (SU(2)\times U(1)_T)$. The internal metric $g_{mn}$ must take the form
\be
g_{mn}(y) \mathd y^m \mathd y^n =  g_{r r}(r) \mathd r^2 + 2 g_{r \psi}(r) \mathd r \mathd\psi +
  g_{\psi \psi}(r) \mathd \psi^2 + e^{2 C(r)} \sum_{a} \mathd u^a \mathd\bar{u}^a\,.
\ee
This metric is not in general Hermitian with respect to the complex structure defined by $(z,u^i)$, but can always be made Hermitian by a suitable coordinate redefinition that alters the complex structure,
\be
z \rightarrow \lambda(r) z \,.
\ee
Using this, the metric may be brought to the form
\be\label{eq:g-flat}
  \mathd s^2 = e^{- 4 B (r)} \left(\mathd r^2 + r^2 \mathd \psi^2\right) + e^{2 C (r)}  \sum_a \mathd u^a \mathd \bar{u}^{a}\,.
\ee
The metric is now manifestly Hermitian, with K\"ahler form
\be\label{eqn:flatkahlerform}
  J = \frac{i}{2}  \Bigl[ e^{- 4 B (r)} \mathd z \wedge \mathd \bar{z} +
  e^{2 C (r)}  \sum_a \mathd u^a \wedge \mathd \bar{u}^a  \Bigr]\,.
\ee
The metric is K\"ahler if and only if $e^{2 C(r)}$ is a constant.

We now consider the general form of $G_3$. Since there are no invariant one-forms pointing along the base (the $\mathd u^a$ and $\mathd \bar{u}^a$ directions), invariant three-forms must have two legs along the base and one along the fiber, and descend from invariant two-forms along the base. These are:
\be
\omega_{1,1}  \equiv \frac{i}{2} \sum_a \mathd u^a \wedge \mathd \bar{u}^a\;\;,\;\;
\omega_{2,0} \equiv z \,\mathd u^1 \wedge \mathd u^2\,.
\ee
Invariant three forms are constructed by wedging $\frac{1}{z} \mathd z$ and its conjugate into $\omega_{1,1}$, $\omega_{2,0}$, and $\omega_{0,2}\equiv\omega_{2,0}^{\star}$. Three-forms built out of $\omega_{1,1}$  are even under $z\rightarrow -z$, whereas those built from $\omega_{2,0}$ and $\omega_{0,2}$ are odd.
$G_3$ then takes the general form:
\be\label{eqn:flatflux}
G_3 = g_{3, 0}  \mathd z \wedge \mathd u^1 \wedge \mathd
  u^2 + g_{2, 1} e^{2i\psi} \mathd \bar{z} \wedge \mathd u^1 \wedge \mathd
  u^2+g_{1, 2} e^{-2 i\psi} \mathd z \wedge \mathd \bar{u}^1 \wedge
  \mathd \bar{u}^2 + g_{0, 3}  \mathd \bar{z} \wedge \mathd
  \bar{u}^1 \wedge \mathd \bar{u}^2\,,
\ee
where the $g_{p,q}$ are complex-valued functions of $r$ only. Comparing with (\ref{eqn:flatkahlerform}), we see that $J \wedge G_3 = 0$, so that $G_3$ is automatically primitive.

The scalars $\alpha$, $A$, and $\tau$ can depend only on $r$. As previously remarked, the $U(1)_{\psi}$ symmetry is broken for solutions with non-vanishing $G_3$. In \S\ref{subsec:various} this breaking
will be identified with the spontaneous breaking of the exact R-symmetry on the gauge theory side.

Although one can obtain an exact supersymmetric solution to the above system by writing down the equations of motion\footnote{For reference, we summarize the equations of motion in Appendix \ref{appendix:eoms}.} and solving them directly, it is far easier to use the conditions for unbroken supersymmetry, which we will present in \S\ref{sec:SUSY} and solve in \S\ref{sec:solution}. First, however, we generalize the preceding ansatz to a Calabi-Yau cone.

\subsection{Seven-branes in the $\mathbb{P}^2$ cone}\label{subsec:P2}

Having proposed an ansatz for the relatively simple geometry near a seven-brane stack, we now extend our analysis to seven-branes wrapping the $\mathbb P^2$ base of the resolved orbifold $\mathbb C^3/\mathbb Z^3$.  We will verify that the complete supergravity ansatz proposed in \S\ref{subsec:flat} emerges upon taking the near-stack limit of our result for the $\mathbb{P}^2$ cone.  This connection provides valuable insight into the near-stack solution, in particular because knowledge of the solution for a compact four-cycle provides a regulator for divergences associated with the noncompact nature of the four-cycle in the near-stack ansatz.\footnote{For example, $\alpha^{\prime}$ corrections induce D3-brane charge on certain compact four-cycles, but the topological information determining  this charge is lost in taking the near-stack limit.}

We develop the ansatz using the $\mathbb{C}^3/\mathbb{Z}_3$ coordinates, and later shift to a different chart appropriate for studying a small neighborhood of the resolved $\mathbb{P}^2$. As before, we wrap  eight D7-branes on the resolved $\mathbb{P}^2$ and orientifold using the involution $z^i \rightarrow -z^i$ combined with $-1 \in SL(2,\mathbb{Z})$. The Calabi-Yau metric has an isometry group $U(3)$ which acts naturally on the $z^i$. However, $z^i \rightarrow -z^i$ lies within the $U(1)$ factor, so nonzero $G_3$ will spontaneously break the $U(1)$, leaving an $SU(3)$ symmetry group.
We now study the ansatz
that arises upon imposing this symmetry group.
Since the orbifold action $\mathbb{Z}_3 \subset SU(3)$, $SU(3)$ singlets are never projected out by the orbifold, and we need not consider $\mathbb{Z}_3$ invariance separately.

\subsubsection{Metric ansatz for the $\mathbb{P}^2$ cone} \label{subsec:P2metric}

The backreacted ten-dimensional metric will be of the form (\ref{eqn:10dmetric}), where the internal metric $g_{mn}$ must be invariant under the $SU(3)$ symmetry.
Since the symmetry group acts transitively on the horizon, we are free to choose a particular point or region on the horizon $S^5/\mathbb{Z}_3$, and we select the $z^1 = z^2 = 0$ plane. The symmetry group within this plane decomposes to $SU(2) \times U (1)$. The metric evaluated in this plane must take the $SU(2) \times U(1)$ invariant form:
\be
  g_{mn}\mathd y^m \mathd y^n  =  \mathfrak{f}_1 (|z^3 |) \left(\bar{z}^3/z^3\right) \mathd z^3 \mathd z^3 + c.c.
      +\mathfrak{f}_2 (|z^3 |) \mathd z^3 \mathd \bar{z}^3
      +\mathfrak{f}_3 (|z^3 |) (\mathd z^1 \mathd \bar{z}^1 + \mathd z^2 \mathd \bar{z}^2)\,.
 \ee
Since $\mathfrak{f}_1$ is complex, the metric depends on four real functions.
We can construct the global form of the metric by combining the invariant
one-forms $\partial \rho^2$,  $\bar{\partial} \rho^2$, and $\sum_i \mathd z^i \mathd \bar{z}^i$:
\be \label{eqn:prep2metric}
  g_{mn}\mathd y^m \mathd y^n = \frac{1}{\rho^2} \mathfrak{f}_1 (\rho) (\partial \rho^2) (\partial \rho^2) + c.c.
  + \frac{1}{\rho^2}  \Biglp  \mathfrak{f}_2 (\rho) -\mathfrak{f}_3 (\rho) \Bigrp  (\partial \rho^2) ( \bar{\partial} \rho^2)
   +\mathfrak{f}_3 (\rho) \sum_i  \mathd z^i \mathd \bar{z}^i \,.
\ee
This reduces to the local form given above at $z^1 = z^2 = 0$, as $\partial
\rho^2 = \sum_i  \bar{z}^i \mathd z^i$. Since $\rho^2$ is $SU(3)$ invariant,
as is the complex structure, the ansatz (\ref{eqn:prep2metric}) is manifestly invariant.

The metric (\ref{eqn:prep2metric}) can always be made Hermitian by a suitable redefinition of the complex structure that
preserves the $SU(3)$ symmetry,
\be
  z^i \rightarrow \lambda (\rho) z^i \,,
\ee
where $\lambda \in \mathbb{C}^{\star}$. One can show that $\mathfrak{f}_1$ can always be set to zero by an appropriate choice of $\lambda$. We define
\be
\mathfrak{f}_2 (\rho) \equiv \rho^4 e^{-4B(\rho)}\;\;,\;\;\mathfrak{f}_3 (\rho) \equiv \frac{1}{\rho^2} e^{2C(\rho)}\,,
\ee
to make the positivity of the metric explicit. Thus,
\be\label{eq:int-P2}
g_{mn}(y) \mathd y^m \mathd y^n = \left(\rho^2 e^{-4B(\rho)}  - \frac{1}{\rho^4} \,e^{2C(\rho)} \right)\sum_{i,j}\bar{z}_i z_j dz^i d \bar{z}^j+ e^{2C(\rho)}\,\frac{1}{\rho^2}\,\sum_i dz^i d \bar{z}^i \,.
\ee
The choice of notation is not accidental; we will see shortly that in the near-stack limit this form reduces to (\ref{eq:g-flat}).

The corresponding K\"ahler form,
\be
  J = \frac{i}{2 \rho^2} \Biglp \rho^4 e^{-4B(\rho)} -\frac{1}{\rho^2} e^{2C(\rho)}\Bigrp  \partial \rho^2 \wedge
  \bar{\partial} \rho^2 + \frac{i}{2 \rho^2} e^{2C(\rho)} \,\partial \bar{\partial} \rho^2\,,
\ee
can be rewritten as $J=e^{-4B}\chi_{1, 1} +e^{2C} \omega_{1, 1}$, where
\be
  \omega_{1, 1} \equiv \frac{i}{2 \rho^2}  \left( \partial \bar{\partial} \rho^2
  - \frac{1}{\rho^2} \partial \rho^2 \wedge \bar{\partial} \rho^2 \right) \;\;,\;\;
  \chi_{1, 1} \equiv \frac{i \rho^2}{2} \partial \rho^2 \wedge \bar{\partial}
  \rho^2\,.
\ee
Note that $\omega_{1,1}$ points along the base, and $\chi_{1,1}$ points along the fiber.

In components, the metric is
\be
  g_{i \bar{j}} = \frac{\rho^4}{2} e^{-4B} P_{i \bar{j}} +
 \frac{1}{2 \rho^2} e^{2C}  \left( \delta_{i \bar{j}} - P_{i \bar{j}} \right) \;\;,\;\; \mathrm{where} \;\;
P_{i \bar{j}} = \frac{1}{\rho^2}  \bar{z}_i z_{\bar{j}} = \delta_{i \bar{k}}
   \delta_{\bar{j} l} P^{\bar{k} l} \;\;,\;\;
z_{\bar i} \equiv \delta_{\bar{i} j} z^j\,,
\ee
and $P^m_n$ is a real rank  one projector satisfying $P^i_j z^j = z^i$.
The determinant of the metric is
\be \label{determinant}
  8 G \equiv 8 \det g_{i \bar{j}} = \frac{8}{3!} \varepsilon^{i j k}
  \varepsilon^{\bar{l}  \bar{m}  \bar{n}} g_{i \bar{l}} g_{j \bar{m}} g_{k
  \bar{n}} = \sqrt{\det g_{m n}} = e^{-4B+4C}\,.
\ee

\subsubsection{K\"ahlerity and the Calabi-Yau metric}

Let us rewrite the metric in terms of the alternate radial coordinate $r \equiv \frac{1}{3} \rho^3$. We find
\be \label{eqn:omegachi_def}
  \omega_{1, 1} = \frac{i}{6 r^2}  \left( \partial \bar{\partial}
  r^2 - \frac{1}{r^2} \partial r^2 \wedge \bar{\partial} r^2
  \right) \;\;,\;\;
  \chi_{1, 1} = \frac{i}{2 r^2} \partial r^2 \wedge \bar{\partial}
  r^2\,.
\ee
Note that $\mathd \omega_{1, 1}= 0$ and $\mathd \chi_{1,1} = -3\,\mathd r^2\wedge \omega_{1,1}$.
The K\"ahler condition now takes a particularly simple form:
\be
\mathd J = \left[ - 3\, e^{-4 B} + \frac{1}{2 r}
   \left(e^{2 C}\right)' \right] \mathd r^2 \wedge \omega_{1, 1} = 0 \,.
\ee
Thus, K\"ahlerity requires $e^{-4 B}= \frac{1}{6 r} \left(e^{2 C}\right)'$, where $f' \equiv \frac{\mathd}{\mathd r} f$. For a K$\ddot{\text{a}}$hler metric, (\ref{determinant}) implies
\be
8 G = \frac{1}{6 r}   \left(e^{2 C}\right)' e^{4 C} = \frac{1}{18
   r}  \left(e^{6 C}\right)' \,.
\ee

As a check, we consider the special case in which the metric is Calabi-Yau. The Ricci form is
\be
   \mathfrak{R}= - i \partial \bar{\partial} \log G = - \frac{i}{2 r}
   \frac{\mathd}{\mathd r}  \left[ \frac{G'}{2 r G} \right]
   \partial r^2 \wedge \bar{\partial} r^2 - \frac{i G'}{2 r G}
   \partial \bar{\partial} r^2 = - \frac{1}{r}  \frac{\mathd}{\mathd
   r}  \left[ \frac{r G'}{2 G} \right] \chi_{1, 1} - \frac{3 r
   G'}{G} \omega_{1, 1}\,.
\ee
Thus, $G' = 0$. We find the solution $e^{6 C} = 9\, \mathfrak{g}_0^3  \left[ r^2 + r_0^2 \right]$, so that
\be\label{eq:resolvedCY}
  e^{2 C} = \mathfrak{g}_0  \left[9\left( r^2 + r_0^2\right) \right]^{1/3} \;\;,\;\;
  e^{-4 B} = \mathfrak{g}_0  \left[9\left( r^2 + r_0^2\right) \right]^{- 2 / 3}\,.
\ee
For $r_0 > 0$ this is the Calabi-Yau metric for the resolved $\mathbb{P}^2$ cone, and for $r_0 = 0$ it describes the singular $\mathbb{P}^2$ cone, for which the metric reduces to the canonical one on $\mathbb{C}^3$ with overall scale $\mathfrak{g}_0$.

\subsubsection{$G_3$ ansatz for the $\mathbb{P}^2$ cone}

We now enumerate the $SU(3)$-invariant $p$-forms of various ranks. The only invariant one-forms are $\partial \rho^2$ and its conjugate, which point along the fiber. In addition to the invariant two-forms $\chi_{1,1}$ and $\omega_{1,1}$ discussed above, pointing along the
fiber and the base, respectively, there exists a complex invariant $(2,0)$ form pointing along the base:
\be
  \omega_{2, 0} \equiv \frac{1}{6} \varepsilon_{i j k} z^i \mathd z^j \wedge
  \mathd z^k\,,
\ee
Including the conjugate $\omega_{0,2} \equiv \omega_{2,0}^{\star}$, this exhausts the list of invariant two-forms. All invariant $p$-forms for $p\ge 3$ can be written as wedge products of invariant one-forms and two-forms. Note that $\omega_{2,0} \wedge \omega_{1,1}=0$, as  both point along the base. These results can be checked by considering the $z^1=z^2=0$ plane as in \S\ref{subsec:P2metric}.

We now consider the most general form of $G_3$
that preserves the $SU(3)$ and is odd under $z^i\rightarrow-z^i$. Since $\omega_{1,1}$ and $\chi_{1,1}$ are even under the involution, whereas $\omega_{2,0}$ and $\omega_{0,2}$ are odd, $G_3$ takes the general form
\be
G_3 = g_{3,0}(r)\, \omega_{3,0}+g_{2,1}(r)\, \omega_{2,1}+g_{1,2}(r)\, \omega_{1,2}+g_{0,3}(r)\, \omega_{0,3}\,,
\ee
where
\be
\omega_{3, 0} \equiv \mathd z^1 \wedge \mathd z^2 \wedge \mathd z^3 =
   \frac{1}{r^2} \partial r^2 \wedge \omega_{2, 0} = \mathd \omega_{2,
   0} \equiv \omega_{0, 3}^{\star}\,,
\ee
\be
\omega_{2, 1} \equiv \frac{1}{r^2}  \bar{\partial} r^2 \wedge \omega_{2,
   0} = \frac{1}{2 \rho^2} \varepsilon_{k l m} z_{\bar{j}} z^k \mathd \bar{z}^{\bar{j}} \wedge
   \mathd z^l \wedge \mathd z^m \equiv \omega_{1, 2}^{\star}\,.
\ee
We immediately find $\omega_{1,1}\wedge G_3=\chi_{1,1}\wedge G_3=0$; therefore $G_3$ is automatically primitive. In components, $G_{i j k} = \varepsilon_{i j k} g_{3, 0}$ and $G_{\bar{i} j k} = \frac{1}{\rho^2} z_{\bar{i}} \varepsilon_{j k l} z^l g_{2, 1}$.

\subsubsection{D3-brane charge}

The seven-brane stack wrapping the $\mathbb{P}^2$ can carry additional charges besides its seven-brane charge, which is fixed by the $-1\in SL(2,\mathbb{Z})$ monodromy. However, since the horizon $S^5/\mathbb{Z}_3$ has vanishing third Betti number, $F_3$ and $H_3$ must be exact, and the stack cannot carry five-brane charge. We now show how to compute the D3-brane charge of the stack.

We define the D3-brane charge enclosed in a region $R$ via the generalized Gauss's law:
\be\label{eq:def-D3}
Q_{\rm{D3}}(R) \equiv -\oint_{\partial R} \tilde{F}_5=(2 \pi)^4 {\alpha^{\prime}}^2 N_{\rm{D3}}\,,
\ee
where D3-branes carry positive charge $2\kappa_{10}^2\, \mu_3=(2 \pi)^4 {\alpha^{\prime}}^2$, and there is a bulk contribution from the three-form fluxes, $Q_{\rm{D3}} = \int F_3\wedge H_3+Q_{\rm{loc}}$. Using (\ref{eqn:F5ansatz}), we obtain:
\be
\tilde{F}_5 = \Omega_4\wedge \mathd \alpha-e^{-8 A} \star_6 \mathd \alpha \,.
\ee
Integrating over the $S^5/\mathbb{Z}_3$ at constant radius and accounting for the $\mathbb{Z}_2$ involution, we find
\be \label{eqn:P2_QD3}
Q_{\rm{D3}}(r) = \oint r e^{4 C-8 A}\, \frac{\mathd \alpha}{\mathd r} \left(\frac{1}{2}\, \omega_{1,1}^2 \wedge \omega \right)
= \frac{\pi^3}{2}\, r e^{4 C-8 A}\, \frac{\mathd \alpha}{\mathd r} \,,
\ee
where $\omega \equiv \frac{1}{i r} (\partial-\bar{\partial}) r$, $\star_6\, \partial r = \frac{1}{2 i} J^2 \wedge \partial r$, and we use the periods
\be \label{eqn:P2periods}
\int_{\mathbb{P}^2} \frac{1}{2}\, \omega_{1,1}^2 = \frac{1}{2}\, \pi^2 \;\;,\;\; \int_{S^5/\mathbb{Z}_3}  \frac{1}{2}\, \omega_{1,1}^2\wedge \omega = \pi^3 \,.
\ee
Since the integrands are closed, these integrals can be computed over any surface in the specified homology class. The periods (\ref{eqn:P2periods}) can also be used to compute the volume of the resolved $\mathbb{P}^2$:
\be \label{eqn:P2_vol}
{\rm{vol}}(\mathbb{P}^2) = \left. \frac{1}{2}\, \pi^2\, e^{4 C(r)} \right|_{r \to 0} \,.
\ee

The D3-brane defined by (\ref{eq:def-D3}) is sourced in the bulk, and therefore not quantized. We define the Page charge~\cite{Page:1984qv} via the flux integral\footnote{As there is more than one way to solve the $\tilde{F}_5$ Bianchi identity, there are inevitable ambiguities in defining the Page charge~\cite{Marolf:2000cb}. This difficulty does not arise in our setup due to the vanishing of the third Betti number for the horizon $\partial R = S^5/\mathbb{Z}_3$, so that all definitions are related by integration by parts.}
\be
Q_{\rm D3}^{\rm Page}(R) \equiv -\oint_{\partial R} \left(\tilde{F}_5 + \frac{1}{2}\, C_2 \wedge H_3 - \frac{1}{2}\, B_2 \wedge F_3\right)\,.
\ee
The integrand is closed in the absence of sources, so the Page charge is not sourced in the bulk. In the absence of local sources coincident with $\partial R$, the Page  charge is invariant under small gauge transformations of $B_2$ and $C_2$. It is not invariant under large gauge transformations unless $F_3$ and $H_3$ are exact when pulled back to $\partial R$. It has been argued~\cite{Marolf:2000cb} that the Page charge is quantized.

One can solve the Bianchi identity for $G_3$ to obtain
\be
\mathcal{A}_2 \equiv \frac{1}{\sqrt{\tau_2}} \left(C_2-\tau B_2\right)=(g_{3,0}-g_{2,1})\, \omega_{2,0}+(g_{0,3}-g_{1,2})\, \omega_{0,2}\,.
\ee
It is then straightforward to compute the Page charge of the stack by the same method as above:
\be \label{eqn:P2_QD3Page}
Q_{\rm D3}^{\rm Page} = \frac{\pi^3}{2} \left(r\, e^{4 C-8 A}\, \frac{\mathd \alpha}{\mathd r}+2\, r^2\, |g_{3,0}-g_{2,1}|^2-2\, r^2\,  |g_{1,2}-g_{0,3}|^2\right)\,,
\ee
where the right-hand side is independent of $r$ as a consequence of the Bianchi identities. Since $F_3$ and $H_3$ are exact, the Page charge is gauge invariant. Henceforward, when we refer to the D3-brane charge of the solution, we mean the Page charge given by (\ref{eqn:P2_QD3Page}).

\subsection{The near-stack limit of the $\mathbb{P}^2$ cone}
\label{subsec:nearstacklimit}

We now apply the near-stack limit (\ref{eqn:nearstacklimit}) to the above ansatz to recover the near-stack ansatz described in \S\ref{subsec:flat} and to fix the
relationship between the fields. As before, we apply the coordinate transformation (\ref{eqn:nearstackchart}), rescale as in (\ref{eqn:nearstacklimit}),
%
and truncate to leading order in $\varepsilon$. Finally, since $\varepsilon$ is a formal expansion parameter, we set $\varepsilon=1$.  We find
\be
\partial r^2  \rightarrow  \partial r_z^2 \;\;,\;\;
\chi_{1,1}  \rightarrow  \frac{i}{2}\, \mathd z \wedge \mathd \bar{z} \;\;,\;\;
\omega_{1,1}  \rightarrow  \frac{i}{2} \left( \sum_a \mathd u^a \wedge \mathd \bar{u}^a\right) \,.
\ee
Thus, in particular
\be
J \rightarrow \frac{i}{2} \Bigl(e^{-4 B} \mathd z \wedge \mathd \bar{z}+e^{2 C} \sum_a \mathd u^a \wedge \mathd \bar{u}^a \Bigr) \,.
\ee
We also find
\be
G_3 \rightarrow g_{3,0} \mathd z\wedge \mathd u^1\wedge \mathd u^2+g_{2,1} \frac{z}{\bar{z}} \mathd \bar{z}\wedge \mathd u^1\wedge \mathd u^2+g_{1,2} \frac{\bar{z}}{z} \mathd z\wedge \mathd \bar{u}^1\wedge \mathd \bar{u}^2+g_{0,3} \mathd \bar{z}\wedge \mathd \bar{u}^1\wedge \mathd \bar{u}^2\,.
\ee
Therefore the quantities $r$, $B$, $C$, and $g_{p,q}$ defined in this section are appropriate generalizations far from the stack of the quantities $r$, $B$, $C$, and $g_{p,q}$ defined in \S\ref{subsec:flat}, and the two systems correspond in the near-stack limit. As an example, applying the near-stack limit to the Calabi-Yau metric for the resolved cone (\ref{eq:resolvedCY}) gives a flat metric with $B$ and $C$ constant. The parameters $r_0$ and $\mathfrak{g}_0$ of the $\mathbb{P}^2$ ansatz can be recovered from
\be \label{eqn:NS_CYmetric}
r_0 = \frac{1}{3}\, e^{2 B_{\rm{ns}}+ C_{\rm{ns}}}\;\;,\;\; \mathfrak{g}_0 = e^{4 (C_{\rm{ns}}-B_{\rm{ns}})/3}\,,
\ee
where $B_{\rm{ns}}$ and $C_{\rm{ns}}$ are the constant values of $B$ and $C$ in the near-stack ansatz.

\section{Supersymmetry Conditions}\label{sec:SUSY}

Our approach to learning about the gauge dynamics of compact seven-branes is to study the allowed supersymmetric backgrounds surrounding the seven-brane stack subject to some symmetry group, in the same spirit that early constructions of supersymmetric extremal black holes in supergravity (cf. e.g.\ \cite{Strominger}) presaged the appearance of D-branes as localized sources in supergravity.

In general, one expects $AdS_4$ vacua from compactifications that are stabilized by gaugino condensation on seven-branes~\cite{Kachru:2003aw}. This is because the superpotential generically develops a vev. If the compactification has finite warped volume, and therefore a finite four-dimensional Newton constant, the superpotential vev generates a negative cosmological constant, leading to an $AdS_4$ compactification. Noncompact solutions may be Minkowski, but we expect such solutions to arise in an appropriate decompactification limit of an $AdS_4$ solution.

We first consider general properties of supersymmetric $AdS_4$ compactifications, and return to the question of noncompactness below. As we will see, supersymmetric compactifications of type IIB supergravity to $AdS_4$ always have $SU(2)$ structure. In our case, the $SU(2)$ structure is dynamic. As dynamic $SU(2)$ structure is less familiar than the more commonly studied strict $SU(3)$ structure, we now review dynamic $SU(2)$ structure in $AdS_4$ compactifications of type IIB supergravity, using tools from the more general field of generalized complex geometry.

\subsection{Review of $SU(2)$ structure and generalized complex geometry}\label{subsec:gcg}

Having argued in \S\ref{subsec:introGCG} that the supergravity solution should admit two spinors $(\eta_+^1, \eta_+^2)$ that define a dynamic $SU(2)$ structure, the next step is to analyze the gravitino and dilatino variations for the supergravity ansatz introduced above. As shown in~\cite{Grana:2005sn}, the supersymmetry analysis simplifies considerably in terms of bispinors. Here we review the basic results from G-structures and generalized complex geometry needed for the rest of the work. For a recent review with further explanations and references, see e.g.~\cite{Koerber:2010bx}.

We work in ten-dimensional Einstein frame, with metric ansatz (\ref{eqn:10dmetric}),
\be
  \mathd s_{10}^2 = e^{2 A (y)} h_{\mu \nu} (x) \mathd x^{\mu} \mathd
  x^{\nu} + e^{- 2 A (y)} g_{m n} (y) \mathd y^m \mathd y^n\,.
\ee
The warp and conformal factors are already made explicit and all the geometric quantities are constructed in terms of the internal metric $g_{mn}$. The four-dimensional metric $h_{\mu\nu}$ is that of AdS, with cosmological constant
\be
\Lambda = -3 |\mu|^2\,.
\ee
The supersymmetry conditions were obtained in~\cite{Grana:2005sn,Grana:2006kf}, in string frame. The conversion from string frame to Einstein frame is done by modifying their warp factor $A_{(S)}$ and pure spinors $\Psi_{(S)}$ as follows:
\be \label{eqn:string_einst_conv}
A_{(S)}= A +\frac{\phi}{4}\;\;,\;\;\Psi_{(S)} = e^{(\phi/4-A) \hat p} \Psi \,,
\ee
where the definition of the operator $\hat p$ is given by
\be
\hat p \, C_p \equiv p\, C_p
\ee
for a $p$-form $C_p$. The rescaling in the pure spinor takes into account that the
Mukai pairing $\langle \Psi, \overline \Psi \rangle$ is normalized by the volume of the internal metric $\mathd^6y \sqrt{\det g_{mn}}$.

As in (\ref{eq:10dspinor}), we decompose the ten-dimensional Majorana-Weyl supersymmetry generators $\epsilon^i$ in Einstein frame,\footnote{Using the conventions of~\cite{Polchinski:1998rr}, $\eta_- = C \eta_+^{\star}$, where $C$ is the charge conjugation matrix.}
\be\label{eq:10dspinor2}
\epsilon^i = \zeta_+ \otimes \eta_+^i + \zeta_- \otimes \eta_-^i \,.
\ee
The internal spinors $\eta^i$ must have equal norms for an $AdS_4$ compactification~\cite{Grana:2006kf}. Preservation of four-dimensional $\mathcal{N}=1$ supersymmetry then imposes the normalization
\be\label{eq:norms}
|\eta_+^1|^2 =|\eta_+^2|^2 =e^A\,,
\ee
up to an arbitrary overall rescaling.

The two internal spinors may be combined into even and odd bispinors,
\be\label{eq:def-bispinors}
\Psi_+ =-8i\, e^{-A}\,\eta_+^1 \otimes \eta_+^{2 \dag}\;\;,\;\;\Psi_- =-8i\,e^{-A}\,\eta_+^1 \otimes \eta_-^{2 \dag}\,,
\ee
where the extra warp factor dependence has been included for normalization purposes. Using the Clifford map, these bispinors are sums of forms of different degrees (polyforms). The supersymmetry conditions of~\cite{Grana:2005sn} then become
\bea\label{eq:SUSY}
\mathd_H \left( e^{(\phi/4-A)\hat p} e^{4A}\,{\rm Re}\,\Psi_+\right)&=& -3 e^{(\phi/4-A)\hat p}  e^{3A-\phi/4}\,{\rm Re}(\bar \mu \Psi_-)+e^{(2A-\phi/2)(3-\hat p)} e^{4A+\phi} \star_6 \lambda(F)\,, \nonumber\\
\mathd_H\left( e^{(\phi/4-A)\hat p} e^{2A-\phi/2}\,{\rm Im}\,\Psi_+\right)&=& 0\,, \nonumber\\
\mathd_H\left( e^{(\phi/4-A)\hat p} e^{3A-\phi/4}\,\Psi_-\right)&=& -2 i \mu\,e^{(\phi/4-A)\hat p} e^{2A-\phi/2}\,{\rm Im}\,\Psi_+\,.
\eea
Here $\mathd_H\equiv \mathd - H \wedge$, and all fluxes are internal, with
\be\label{eq:conventionF}
F \equiv F_1 + \t{F}_3 + \t{F}_5^{({\rm int})}\;\;,\;\;\lambda(A_p)=(-1)^{p(p-1)/2} A_p\,.
\ee
When supplemented by the $p$-form Bianchi identities, the above supersymmetry conditions imply all of the supergravity equations of motion (which we will verify explicitly in our examples).\footnote{In the presence of sources, one must also impose calibration conditions on the sources; cf.~\cite{Koerber:2007hd}.}

Now, let us consider the geometric properties of manifolds with $SU(2)$ structure. It is convenient to introduce two globally defined orthonormal spinors $\eta_+$ and $\chi_+$. They are related by a vector $\Theta_m$,
\be
\chi_+ = \frac{1}{2} \Theta^m \gamma_m \eta_-\,,
\ee
where $|\Theta|^2=2$. An $SU(2)$ structure is then characterized by the following invariant forms:
\be\label{eq:SU2tensors}
\Theta_m = \eta_-^\dag \gamma_m \chi_+ \;\;,\;\;
(J_2)_{mn} = - \frac{i}{2}\Bigl( \eta_+^\dag \gamma_{mn} \eta_+ - \chi_+^\dag \gamma_{mn} \chi_+\Bigr) \;\;,\;\;
(\Omega_2)_{m n} = i \chi_+^{\dag} \gamma_{m n} \eta_+ \,.
\ee
These satisfy
\be
J_2 \wedge \Omega_2 =  \Omega_2 \wedge \Omega_2 = 0\;\;,\;\;
J_2 \wedge J_2 = \frac{1}{2} \Omega_2 \wedge \overline \Omega_2 \;\;,\;\;
\imath_\Theta \Omega_2 =  \imath_\Theta J_2 =0\,.
\ee
Algebraically, the tangent bundle has a product structure, where $\Omega_2$ and $J_2$ may be thought of as the holomorphic two-form and K\"ahler form for a complex-dimension two subspace of the tangent bundle, and $\Theta$ and $J_1 \equiv \frac{i}{2} \Theta \wedge \bar{\Theta}$ as the holomorphic one-form and K\"ahler form for the complex-dimension one complement. However, this product structure is typically not integrable, and the manifold need not be a direct product. Instead, we will think of the manifold as a line bundle with $J_2$ and $\Omega_2$ pointing along the base and $\Theta$ and $\bar{\Theta}$ pointing along the fiber. This structure will turn out to be integrable in our examples, but this is not guaranteed in general.

The $SU(2)$ structure can be viewed locally as the intersection of two $SU(3)$ structures, each associated to one of the spinors. In particular, the $SU(3)$ structure from $\eta_+$ is defined by
\be\label{eq:SU3tensors}
J_{mn} = -i \eta_+^\dag \gamma_{mn} \eta_+\;\;,\;\;\Omega_{mnr} = i \eta_-^\dag \gamma_{mnr} \eta_+\,.
\ee
The forms (\ref{eq:SU2tensors}) and (\ref{eq:SU3tensors}) are related by
\be
J = J_2+ \frac{i}{2} \Theta \wedge \bar{\Theta}\;\;,\;\;\Omega = \Theta \wedge \Omega_2\,.
\ee
There are different ways of writing the spinors $\eta^i_+$ in terms of $(\eta_+, \chi_+)$. In geometries with orientifold planes it is most convenient to average $\eta_+\propto (\eta^1_+ + i e^{i \vartheta} \eta^2_+)$, where $i e^{i \vartheta}$ is the relative phase between the two spinors~\cite{Minasian:2006hv,Andriot:2008va}. Thus we take
\be\label{eq:eta-basis}
\eta_+^1 = i\, e^{i\, \vartheta/2}\, e^{A/2} \left(\cos \frac{\varphi}{2} \,\eta_+ +\sin \frac{\varphi}{2}\, \chi_+ \right) \;\;,\;\;
\eta_+^2 = e^{-i\, \vartheta/2}\, e^{A/2} \left(\cos \frac{\varphi}{2} \,\eta_+ - \sin \frac{\varphi}{2}\, \chi_+ \right) \,.
\ee
The warp factor is fixed by the normalization (\ref{eq:norms}), and $\vartheta$ and $\varphi$ parameterize the angle between the spinors,
\be
e^{-A}\, \eta_+^{2\, \dag} \eta_+^1=i\, e^{i \vartheta} \cos \varphi\,.
\ee

Using this, the bispinors may be expressed  in terms of the $SU(2)$ forms (\ref{eq:SU2tensors}), yielding
\bea\label{eq:SU2-bispinor}
  \Psi_+ & = & e^{i \vartheta} e^{\frac{1}{2} \Theta \wedge \bar{\Theta}}  \left[ \cos
  \varphi \left( 1 - \frac{1}{2} J_2^2 \right) + \sin \varphi\, \mathrm{Im}\,
  \Omega_2 - i J_2 \right]\,, \nonumber \\
  \Psi_- & = & \Theta \wedge \left[ \sin \varphi \left( 1 - \frac{1}{2} J_2^2 \right)
  - \cos \varphi\, \mathrm{Im}\, \Omega_2 + i\, \mathrm{Re}\, \Omega_2
  \right]\,.
\eea
It is straightforward to check by substituting this result into (\ref{eq:SUSY}) that we must take $e^{i\, \vartheta}=\pm 1$ for $AdS_4$ solutions, where the extra sign can be absorbed by redefinitions. Thus, we take $\vartheta=0$ without loss of generality.

The supersymmetry structure is characterized by the angle $\varphi$, which in turn determines the types of $\Psi_+$ and $\Psi_-$.\footnote{Here the `type' of a polyform $\Psi$ refers to the rank of the lowest-rank non-zero component of $\Psi$, as in~\cite{Gualtieri}. In what follows, type $(m,n)$ refers to $\Psi_+$ ($\Psi_-$) of type $m$ ($n$).} For static $SU(2)$ structure (type $(2,1)$), $\varphi = \pi/2$ and the two internal spinors are everywhere orthogonal. For strict $SU(3)$ structure (type $(0,3)$), $\varphi=0$ and the spinors are everywhere parallel; the polyforms simplify to $\Psi_+ =e^{i\,\vartheta} e^{-iJ}$ and $\Psi_-= i\, \Omega_3$. For intermediate $SU(2)$ structure (type $(0,1)$), $0 < \varphi < \pi/2$, and the spinors are neither parallel nor orthogonal.

If $\varphi$ varies along the internal manifold, the $SU(2)$ structure is said to be dynamic. Dynamic $SU(2)$ structure can be `type-changing' if $\varphi = 0$ or $\varphi = \pi/2$ on some locus. Our solution will turn out to be dynamic; for certain values of the parameters, it is also type-changing with a $\varphi = \pi/2$ locus.

We now impose the orientifold projection
$\mc O =\Omega_p (-1)^{F_L} \sigma$, where $\sigma$ is the involution, $F_L$ is the number of left-moving fermions, and $\Omega_p$ is the worldsheet parity. The involution on the pure spinors $(\Psi_+, \Psi_-)$ should reduce to the known result
\be
\sigma(J) = J\;\;,\;\;\sigma(\Omega) =- \Omega\,,
\ee
for an O3/O7, so that we have~\cite{Grana:2006kf},
\be\label{eq:involution-bispinor}
\sigma(\Psi_+) =  \lambda (\overline \Psi_+)\;\;,\;\;\sigma(\Psi_-)= \lambda(\Psi_-)\,,
\ee
with $\lambda$ defined in (\ref{eq:conventionF}). Applying (\ref{eq:involution-bispinor}) to (\ref{eq:SU2-bispinor}) gives
\be \label{eq:SU2-involution}
\sigma (J_2) = J_2\;\;,\;\;\sigma(\Omega_2)= -\Omega_2\;\;,\;\; \sigma(\Theta) = \Theta \;\;,\;\; \sigma(\varphi) = \varphi\,.
\ee
In other words, in the basis (\ref{eq:eta-basis}) the orientifold action is realized as an explicit $\pi$ rotation in the $({\rm Re} \,\Omega_2, {\rm Im}\, \Omega_2)$ `plane' of the space $(J_2, {\rm Re} \,\Omega_2, {\rm Im}\, \Omega_2)$~\cite{Andriot:2008va}.

For strict $SU(3)$ structure compactifications, we must take the holomorphic three-form $\Omega$ to carry R-charge $+2$~\cite{Klebanov:1998hh}. Generalizing this, we see that $\Psi_-$ should carry R charge $+2$. Thus, $\Theta$ and $\mu$ carry R-charge $+2$, while $\Omega_2$ and $J_2$ are invariant.

Substituting (\ref{eq:SU2-bispinor}) into (\ref{eq:SUSY}), we obtain the supersymmetry conditions for compactifying type IIB supergravity to $AdS_4$. As remarked above, supersymmetric compactification to $AdS_4$ requires $\eta_+^{1\, \dag} \eta_+^1=\eta_+^{2\, \dag} \eta_+^2$ and $\mathrm{Re}\,\eta_+^{2\, \dag} \eta_+^1=0$. We refer to Minkowski ($\mu=0$) solutions that satisfy these conditions as `AdS-like.' The set of `AdS-like' solutions may be thought of as the closure of the set of AdS solutions, since a solution
that arises upon taking a limit in parameter space of an AdS solution must still satisfy $\eta_+^{1\, \dag} \eta_+^1=\eta_+^{2\, \dag} \eta_+^2$ and $\mathrm{Re}\,\eta_+^{2\, \dag} \eta_+^1=0$, provided that $\eta^1$ and $\eta^2$ vary continuously in this limit. Thus, we expect that noncompact solutions with gaugino condensation will be AdS-like. AdS-like solutions with strict $SU(3)$ structure are the well known type B supersymmetric solutions which arise from F-theory compactifications.

The supersymmetry conditions for AdS and AdS-like compactifications can be rewritten in an $SL(2,\mathbb{R})$ covariant form. In \S\ref{subsec:SL2Rcovar}, we briefly review $SL(2,\mathbb{R})$ covariant supergravity. In \S\ref{subsec:SL2Rsusy}, we present the covariant supersymmetry conditions, deferring a detailed derivation to \cite{Benforthcoming}, and discuss their implications.

\subsection{$SL(2,\mathbb{R})$ covariant supergravity} \label{subsec:SL2Rcovar}

It is well known that the action (\ref{eqn:SUGRAaction}) is invariant under $SL(2,\mathbb{R})$ transformations, where $\tau = C_0 + i e^{- \phi}$ transforms as $\tau \rightarrow \frac{a \tau + b}{c \tau + d}$, $F_3$ and $H_3$ mix as:
\bea
  \left(\begin{array}{c}
    F_3\\
    H_3
  \end{array}\right) & \rightarrow & \left(\begin{array}{cc}
    a & b\\
    c & d
  \end{array}\right) \left(\begin{array}{c}
    F_3\\
    H_3
  \end{array}\right)\,,
\eea
with $a d - b c = 1$, and the metric and $C_4$ are invariant. Accounting for
brane sources as required by string theory, this $SL(2, \mathbb{R})$
breaks to the discrete subgroup $SL(2, \mathbb{Z})$, but this
subgroup is gauged: monodromies are allowed, and indeed charged
seven-branes carry $SL(2, \mathbb{Z})$ monodromies.

Since we are interested in studying nonperturbative effects on seven-branes,
an approach that
makes $SL(2, \mathbb{Z})$ (and indeed $SL(2, \mathbb{R})$) invariance manifest is indispensable (cf. e.g.~\cite{Cederwall:1997ts}). It will be convenient to work with $\frac{1}{\tau_2}\, \mathd \tau$ and the complex field strength and potential
\be
G_3 \equiv \frac{1}{\sqrt{\tau_2}}\, (F_3 - \tau H_3)\;\;,\;\;\mathcal{A}_2 \equiv \frac{1}{\sqrt{\tau_2}}\, (C_2 - \tau B_2)\,.
\ee
These transform by a
$\tau$-dependent phase under $SL(2, \mathbb{R})$:
\be
  G_3 \rightarrow \frac{|c \tau + d|}{c \tau + d}\, G_3 \;\;,\;\;
  \frac{1}{\tau_2}\, \mathd \tau \rightarrow \left( \frac{|c \tau + d|}{c \tau + d} \right)^2 \frac{1}{\tau_2}\, \mathd \tau\,.
\ee
These are both examples of a more general transformation law,
\be
\Omega \rightarrow \left( \frac{|c \tau + d|}{c \tau + d} \right)^{2 Q} \Omega\,,
\ee
where $Q$ is the `S-charge' of $\Omega$. In this language, $G_3$ and
$\mathcal{A}_2$ carry charge $+ 1 / 2$ and $\frac{1}{\tau_2}\, \mathd \tau$ carries charge $+ 1$.

Since
the $\tau$-dependent phase is in general nonconstant, the derivative of an
S-covariant quantity is not covariant. We introduce a covariant derivative
$\partial_M \rightarrow D_M = \partial_M + i Q K_M$ where $K_M$ is a one-form
connection. One can check that $K_M=\frac{1}{\tau_2}\, \partial_M \tau_1$ transforms appropriately, where $\tau_1 = {\rm Re}\, \tau$. Thus, we define the covariant exterior derivative:
\be
  \mathD\, \Omega \equiv \mathd\, \Omega + i\, Q\, \frac{1}{\tau_2}\, \mathd
  \tau_1 \wedge \Omega\,.
\ee
It is easy to check that $\mathD$ is not nilpotent: $\mathD^2 = \frac{i Q}{\tau_2^2}\, \mathd \tau_1 \wedge \mathd \tau_2$.
We define the modified covariant exterior derivative for $\Omega$
of charge $+1/2$:
\be
  \mathD_{\pm}\, \Omega \equiv \mathD\, \Omega \pm \frac{i}{2\, \tau_2}\,\mathd \tau \wedge \Omega^{\star}\,.
\ee
Similarly, for $\Omega$ of
charge $- 1 / 2$, we define
$\mathD_{\pm}\, \Omega \equiv \mathD\, \Omega \pm \frac{i}{2\, \tau_2}\,\mathd \bar{\tau} \wedge
  \Omega^{\star} = \left( \mathD_{\mp}\, \Omega^{\star} \right)^{\star}$.
One can check that $\mathD_+$ and $\mathD_-$ are both nilpotent.{\footnote{Note that
$\mathD_{\pm}$ are not $\mathbb{C}$-linear ($i \mathD_+ =\mathD_- i$) and do not always obey the product rule.}} The $G_3$ Bianchi identity now
takes the simple form $\mathD_-\, G_3  =  0$,
with the local solution
$G_3 = \mathD_-\, \mathcal{A}_2$.

Expressed in terms of $\tau$ and $G_3$, the type IIB supergravity action (\ref{eqn:SUGRAaction}) becomes
\be \label{eqn:covarSUGRA}
  S = \frac{1}{2 \kappa_{10}^2} \int \mathd^{10} x \sqrt{- g^{(10)}}  \left[ R -
  \frac{1}{2}  \left( \frac{1}{\tau_2^2}\, |\mathd \tau|^2 + |G_3 |^2 + \frac{1}{2} | \tilde{F}_5 |^2
  \right) \right] - \frac{i}{8 \kappa_{10}^2} \int C_4 \wedge G_3 \wedge
  G_3^{\star}\,,
\ee
where, in terms of $G_3$ and $\mathcal{A}_2$, $\tilde{F}_5$ takes the local
form
\be
 \tilde{F}_5 = \mathd C_4 + \left( \frac{i}{4} \mathcal{A}_2 \wedge
  G_3^{\star} + c.c. \right)\,,
\ee
with the Bianchi identity
$  \mathd\, \tilde{F}_5 = \frac{i}{2}\, G_3 \wedge G_3^{\star}\,.$
One can then rewrite the equations of motion as:\footnote{Here $|F_p |^2_{M N} \equiv \frac{1}{2 (p - 1) !}  \left( F_p \right)^{M_1 \ldots M_{p - 1}}{}_M  \left( F^{\star}_p \right)_{M_1 \ldots M_{p - 1} N} + c.c.$, so that $g^{M N} |F_p|^2_{M N} = p\, |F_p|^2$.}
\be
  \mathD_- \star_{10} G_3^{\star} = i\, G_3^{\star} \wedge \tilde{F}_5 \;\;,\;\;
  \mathD \star_{10} \Bigl(  \frac{1}{\tau_2}\,\mathd \bar{\tau} \Bigr) = \frac{i}{2}\, G_3^{\star} \wedge \star_{10} G_3^{\star}\,,
\ee
\be
  R_{M N} = \frac{1}{4\, \tau_2^2} \left(\nabla_M \tau\, \nabla_N \bar{\tau}+c.c.\right) + \frac{1}{2}  \left[ |G_3 |^2_{M
  N} - \frac{1}{4}\, |G_3 |^2\, g^{(10)}_{M N} \right] + \frac{1}{4}\, | \tilde{F}_5 |^2_{M
  N}\,,
\ee
and in addition one must impose the self-duality constraint $\tilde{F}_5 = \star_{10} \tilde{F}_5$. The action and equations of motion are now manifestly covariant under $SL(2,\mathbb{R})$ transformations.\footnote{The full supersymmetric action is also invariant~\cite{Howe:1983sra}.}

Consider the additional $\mathbb{Z}_2$ symmetry of type IIB supergravity under which all of the RR fields are reversed (i.e.\
$C_p \rightarrow - C_p$ and $F_p \rightarrow - F_p$). This is an exact (gauged) symmetry of string theory which we denote $\mathbb{Z}_2^{(RR)}$, corresponding for instance to the involution associated with the O9 plane of the type I theory composed with the $- 1 \in SL(2, \mathbb{Z})$. In the S-covariant language developed above, $\mathbb{Z}_2^{(RR)}$ takes the form:
\be \label{eqn:RRparity}
  G_3 \rightarrow - G_3^{\star} \;\;,\;\;
  \tau \rightarrow - \tau^{\star} \;\;,\;\;
  \tilde{F}_5 \rightarrow - \tilde{F}_5 \,.
\ee
If $\Omega$ carries S-charge $Q$ and transforms as $\Omega \rightarrow - \Omega^{\star}$ under $\mathbb{Z}_2^{(RR)}$, then $\mathD\, \Omega \rightarrow - (\mathD\, \Omega)^{\star}$. For $Q=\pm 1/2$, $\mathD_{\pm}\, \Omega \rightarrow - (\mathD_{\pm}\, \Omega)^{\star}$. Viewing
$\mathbb{Z}_2^{(RR)}$ as $\mathrm{diag}(-1,1) \in SL_{\pm}(2,\mathbb{R})$, it combines with the $SL(2,\mathbb{R})$ invariance discussed above to generate the classical symmetry group $SL_{\pm}(2,\mathbb{R})$, of which the subgroup $SL_{\pm} (2,\mathbb{Z})$ is an exact (gauged) symmetry of string theory.

\subsection{$SL(2,\mathbb{R})$ covariant supersymmetry conditions for $AdS_4$ compactifications} \label{subsec:SL2Rsusy}

Henceforward, we work with the almost complex structure defined by the holomorphic three-form $\Omega \equiv \Omega_2 \wedge \Theta$ with associated K\"ahler form $J = J_1+J_2$, where $J_1 \equiv (i/2) \Theta \wedge \bar{\Theta}$. Consistent with (\ref{eq:SU2-involution}), we assign $\Omega_2$ an S-charge of $-1/2$ and take $J_2$, $\Theta$, and $\varphi$ to be invariant. $G_3$ can be decomposed into pieces with zero, one, or two legs along the `fiber' ($\Theta, \bar{\Theta}$) directions. The supersymmetry conditions directly imply $G_3\wedge \Theta \wedge J_2=G_3 \wedge \bar{\Theta} \wedge J_2=0$. Thus, a general decomposition takes the form:
\be \label{eqn:SUSY_G3ansatz}
  G_3 = J_1 \wedge \mathcal{G}_1 + J_2 \wedge \mathfrak{G}_1 + \mathfrak{g}_{3, 0}\,
  \Omega_2 \wedge \Theta + \mathfrak{g}_{2, 1}\, \Omega_2 \wedge \bar{\Theta} + \mathfrak{g}_{1, 2}\,
  \bar{\Omega}_2 \wedge \Theta + \mathfrak{g}_{0, 3}\,  \bar{\Omega}_2 \wedge \bar{\Theta}
  +\mathcal{G}_{1, 1} \wedge \Theta +\mathfrak{G}_{1, 1} \wedge \bar{\Theta}\,,
\ee
where $\mathcal{G}_1$ and $\mathfrak{G}_1$ are complex one-forms pointing along the base,
the $\mathfrak{g}_{p,q}$ are complex scalars, and $\mathcal{G}_{1,1}$ and $\mathfrak{G}_{1,1}$ are complex primitive (1,1) forms pointing along the base. We also decompose the gradient into fiber and base directions:
\be
  \mathd f = \left. \left( \partial_{\Theta} f \right) \Theta +
  (\bar{\partial}_{\Theta} f \right)  \bar{\Theta} + \mathd_{\Pi} f \,.
\ee

Applying these decompositions to the supersymmetry conditions and simplifying, we obtain:
\be \label{eqn:SUSY_ThetaJ}
  \mathd \left[ e^{4 A} c_{\varphi} \right] = - 3\, e^{2 A}\, s_{\varphi}
  \tmop{Re}\, (\bar{\mu}\, \Theta) + \mathd \alpha \;\;,\;\;
  \mathd \left[ e^{2 A} s_{\varphi} \Theta \right] = 2 i \mu \left(
  c_{\varphi} J_1 + J_2 \right) \;\;,\;\;
  \mathd \left[ c_{\varphi} J_1 + J_2 \right] = 0 \,,
\ee
\be
\mathD_+  \left[ e^{2 A} s_{\varphi} \Omega_2 \right] = c_{\varphi} e^{4 A}
  G_3^{\star} - i e^{4 A} \star G_3^{\star} + \frac{3 \bar{\mu}}{2} (1 + c_{\varphi})
  \Omega_2 \wedge \Theta - \frac{3 \mu}{2} (1 - c_{\varphi}) \Omega_2 \wedge
  \bar{\Theta}\,,  \label{eqn:SUSY_Omega}
\ee
\bea
  \mathfrak{g}_{3, 0} =  -\frac{1 - c_{\varphi}}{2 s_{\varphi}}\,e^{-2 A} \frac{i}{\tau_2} \partial_{\Theta} \tau & \;\;,\;\; &
  \mathfrak{g}_{2, 1} =  \frac{1 + c_{\varphi}}{2 s_{\varphi}}\, e^{-2 A} \frac{i}{\tau_2}  \bar{\partial}_{\Theta} \tau \,, \label{eqn:SUSY_g30_g21}\\
  \mathfrak{g}_{1, 2} = \bar{\mu}\, e^{-4 A} + \frac{1 + c_{\varphi}}{2 s_{\varphi}}\, e^{- 6 A}  \partial_{\Theta}  \Phi_- & \;\;,\;\; &
  \mathfrak{g}_{0, 3} = \mu\,e^{-4 A} - \frac{1 - c_{\varphi}}{2 s_{\varphi}}\, e^{- 6 A}  \bar{\partial}_{\Theta}  \Phi_+ \, \label{eqn:SUSY_g12_g03}
\eea
\be
 \mathcal{G}_{1, 0} \wedge J_2 = - s_{\varphi}^{- 1} e^{- 2 A}  \left(
  \frac{i}{\tau_2}\,  \mathd_{\Pi}\, \tau \right) \wedge \Omega_2 \;\;,\;\;
  \mathfrak{G}_{1, 0} = -c_{\varphi} \mathcal{G}_{1, 0} \,, \label{eqn:SUSY_calfrakG10}
\ee
\bea
  \mathcal{G}_{0, 1} \wedge J_2 & =  &- s_{\varphi}^{- 1} e^{- 6 A}  \left(
  \mathd_{\Pi}  \left[ e^{4 A} \right] - c_{\varphi} \mathd_{\Pi} \alpha
  \right) \wedge \bar{\Omega}_2 \,,\label{eqn:SUSY_calG01} \\
  \mathfrak{G}_{0, 1} \wedge J_2 & = & s_{\varphi}^{- 1} e^{- 6 A}  \left(
  c_{\varphi} \mathd_{\Pi}  \left[ e^{4 A} \right] - \mathd_{\Pi} \alpha
  \right) \wedge \bar{\Omega}_2 \,,\label{eqn:SUSY_frakG01}
\eea
where $c_{\varphi}$ and $s_{\varphi}$ are shorthand for $\cos \varphi$ and $\sin \varphi$, and
\be
\Phi_{\pm} \equiv e^{4 A} \pm \alpha\,.
\ee
By referring to the charge assignments of Table~\ref{table:Scharge}, one can verify that the above equations are manifestly invariant under $SL(2,\mathbb{R})$, and also under a $U(1)_R$ symmetry.

\begin{table}[htdp]
\begin{center}
  \begin{tabular}{|c|cc|c|cc|}
  \hline
    & S-charge & R-charge &  & S-charge & R-charge \\ \hline
    $G_3, \bar{\Omega}_2$ & $+ 1 / 2$ & $0$ & $\mathfrak{g}_{1, 2}$ & $0$ & $-
    2$\\
    $\delta \tau / \tau_2$ & $+ 1$ & $0$ & $\mathfrak{g}_{0,
    3}$ & $0$ & $+ 2$\\
    $\mu, \Theta, \bar{\partial}_{\Theta}$ & $0$ & $+ 2$ & $\mathcal{G}_1,
    \mathfrak{G}_1$ & $+ 1 / 2$ & $0$\\
    $\mathfrak{g}_{3, 0}$ & $+ 1$ & $- 2$ & $\mathcal{G}_{1, 1}$ & $+ 1 / 2$ &
    $- 2$\\
    $\mathfrak{g}_{2, 1}$ & $+ 1$ & $+ 2$ & $\mathfrak{G}_{1, 1}$ & $+ 1 / 2$
    & $+ 2$ \\ \hline
  \end{tabular}
\end{center}
\caption{The S-charge and R-charge of various fields; $\delta \tau$ represents any differential of $\tau$.}
\label{table:Scharge}
\end{table}

In fact, the supersymmetry conditions (\ref{eqn:SUSY_ThetaJ} -- \ref{eqn:SUSY_frakG01})
are covariant under the full $SL_{\pm}(2,\mathbb{R})$ classical symmetry group of type IIB supergravity,
which is generated by $SL(2,\mathbb{R})$ transformations and by $\mathrm{diag}(-1,1) \in SL_{\pm}(2,\mathbb{Z})$.
The latter transformation, (\ref{eqn:RRparity}), takes $G_3 \rightarrow - G_3^{\star}$, $\tau \rightarrow - \tau^{\star}$, and $\alpha \rightarrow - \alpha$,
along with $\Psi_{\pm} \rightarrow -\Psi_{\pm}$, so that
\be
\Omega_2 \rightarrow -\Omega_2^{\star}  \;\;,\;\;
\varphi \rightarrow \varphi+\pi  \;\;,\;\;
J_2 \rightarrow -J_2 \,,
\ee
with the appropriate transformations on the components of $G_3$. In addition, the supersymmetry conditions possess the $\mathbb{Z}_2$ symmetry
\be
\Theta \rightarrow -\Theta \;\;,\;\; \varphi \rightarrow -\varphi \;\;,\;\; \Omega_2 \rightarrow -\Omega_2 \,,
\ee
under which the polyforms $\Psi_+$ and $\Psi_-$ are invariant.

Regular
AdS
solutions that
satisfy these conditions cannot be type changing with an $SU(3)$ structure locus. To show this, we expand out the derivative in the middle equation of (\ref{eqn:SUSY_ThetaJ}):
\be
  c_{\varphi}\, e^{2 A} \mathd \varphi \wedge \Theta + s_{\varphi}\, \mathd \left[
  e^{2 A} \Theta \right] = 2 i \mu \left( c_{\varphi} J_1 + J_2 \right)\,.
\ee
Suppose that $\varphi = 0$ at
some regular point of the solution. Thus,
\be
  e^{2 A} \mathd \varphi \wedge \Theta = 2 i \mu \left( J_1 + J_2 \right)\,.
\ee
For AdS solutions ($\mu \neq 0$), this would imply $J_2 = 0$, which contradicts the assumption of regularity.
The same considerations apply for $\varphi = \pi$ loci.
Thus, a supersymmetric AdS solution that is regular everywhere must have global $SU(2)$ structure, i.e.\ nowhere vanishing $\sin \varphi$, whereas an AdS solution with local singularities must have $SU(2)$ structure away from these singularities. Smooth type-changing between $SU(2)$ and $SU(3)$ structures is not obviously forbidden for AdS-like solutions with $\mu = 0$, though no examples of this are known.

The supersymmetry conditions for type B solutions are well known, and are easily derived from (\ref{eq:SUSY}): $\tau$ must be holomorphic, $J$ must be closed, $\Phi_-$ must vanish, $G_3$ must be primitive and of Hodge type $(2,1)$, and $\mathd (e^{\phi/2}\, \Omega)=0$, which can be restated covariantly as
 $\mathD \Omega = 0$. Unlike type B solutions, AdS and AdS-like $SU(2)$ structure solutions need not be K\"ahler or even complex. Taking the $(1,2)$ component of (\ref{eqn:SUSY_Omega}) and applying (\ref{eqn:SUSY_g30_g21}, \ref{eqn:SUSY_calfrakG10}) we obtain
\be
  \left[\mathd\, \Omega_2\right]_{1,2} =  \frac{1}{2}\, s_{\varphi}\, e^{2 A} J_2 \wedge \left( \mathcal{G}_{1, 0} \right)^{\star}
  - \frac{1 - c_{\varphi}}{s_{\varphi}}\, e^{2 A}\,  \bar{\Theta} \wedge \left( \mathcal{G}_{1, 1} \right)^{\star}\,.
\ee
Applying (\ref{eqn:SUSY_ThetaJ}), we find $\left[\mathd\, \Omega\right]_{2,2} = \left[\mathd\, \Omega_2\right]_{1,2} \wedge \Theta$. Thus, the almost complex structure associated with $\Omega = \Omega_2 \wedge \Theta$ is integrable if and only if $\mathcal{G}_{1,0}$ and $\mathcal{G}_{1,1}$ both vanish. Similarly, the K\"ahler form $J$ is not in general closed. Applying (\ref{eqn:SUSY_ThetaJ}), we find
\be \label{eqn:SUSY_dJ}
  \mathd J = - 2\, \mu\, e^{- 2 A}\, \frac{1 - c_{\varphi}}{s_{\varphi}} J_2
  \wedge \tmop{Re} \Theta -e^{-4 A} \left(\frac{1-c_{\varphi}}{1+c_{\varphi}}\right) \mathd\, \Phi_+ \wedge J_1\,.
\ee
Adding (\ref{eqn:SUSY_calG01}) and (\ref{eqn:SUSY_frakG01}), we see that $\mathd_{\Pi}\, \Phi_+$ vanishes if and only if $\mathcal{G}_{0,1}+\mathfrak{G}_{0,1}=0$. Thus, $J$ is closed if and only if the solution is Minkowski with $\mathcal{G}_{0,1}=-\mathfrak{G}_{0,1}$. Using (\ref{eqn:SUSY_calfrakG10}), the above conditions on $\mathcal{G}_{0,1}$ and on the pair $\mathcal{G}_{0,1}$ and $\mathfrak{G}_{0,1}$ can be efficiently restated as the requirement that $G_{2,1}$ and $G_{1,2}$ be primitive, respectively.

In~\cite{Koerber:2007xk} it is argued that the appropriate generalization of the Gukov-Vafa-Witten on-shell flux superpotential~\cite{Gukov:1999ya} to generalized complex geometry solutions is the on-shell superpotential
\be \label{eqn:KMsuperpotential}
W = \frac{1}{4\, \kappa_{10}^2} \int \left(e^{3 A_{(S)} - \phi}\, \Psi_-^{(S)}\right)\wedge \lambda(F)\,,
\ee
where $A_{(S)}$ and $\Psi_-^{(S)}$ are related to the Einstein frame quantities $A$ and $\Psi_-$ by (\ref{eqn:string_einst_conv}), $\lambda$ is defined in (\ref{eq:conventionF}), and the integral is over the compact manifold. Supersymmetric AdS solutions correspond to a nonvanishing superpotential vev.
Using the supersymmetry conditions (\ref{eqn:SUSY_ThetaJ} -- \ref{eqn:SUSY_frakG01}), we can evaluate (\ref{eqn:KMsuperpotential}) explicitly. We find
\be \label{eqn:onshell_superpotential}
W = -\frac{\mu}{\kappa_{10}^2} \int \mathd^6 y\, \sqrt{g}\, e^{- 4 A} =  -\frac{\mu}{\kappa_4^2}\,,
\ee
in exact agreement with the four-dimensional supergravity result $\Lambda = -3\, \kappa_{4}^4\, |W|^2$, where $\kappa_4^2$ is the four-dimensional Newton constant. The limit of rigid supersymmetry, $\kappa_4^2 \to 0$, is more subtle, as the integrand of (\ref{eqn:KMsuperpotential}) vanishes on-shell, but the integral is taken over an infinite volume.

\section{Supergravity Solution in the Near-stack Region}\label{sec:solution}

In this section we search for supersymmetric solutions to the ansatz described in \S\ref{subsec:flat}. First, as a warmup we describe type B solutions to this ansatz. The axiodilaton must be holomorphic, but can only depend on the real coordinate $r$, so it is a constant, $\tau = C_0+\frac{i}{g_s}$. Thus, the solution is conformally Calabi-Yau, with $B$ and $C$ constant, consistent with the near-stack limit of the metric (\ref{eq:resolvedCY}). The only nonvanishing component of $G_3$ is $G_{2,1}$, which must be closed by the $G_3$ Bianchi identity; thus, $g_{2,1}=e^{2 C}\, k_{2,1}/r^2$. Finally, we take $\alpha=e^{4 A}=\Phi_+/2$, and solve the $\tilde{F}_5$ Bianchi identity, (\ref{eqn:alphaeom}), which gives:
\be
\frac{1}{r} \frac{\mathd}{\mathd r} \left(r\, \left(\Phi_+^{-1}\right)' \right) = -2\, e^{-4 C}\, |g_{2,1}|^2 \,,
\ee
where primes denote derivatives with respect to $r$. We integrate to obtain $\Phi_+^{-1}=k_0+k_1 \log r-\frac{1}{2 r^2} |k_{2,1}|^2$. Comparing with (\ref{eqn:P2_QD3Page}), we see that $k_1$ is related to the
D3-brane charge:
\be
Q_{\rm{D3}}=-\pi^3\, r\, e^{4 C}\, (\Phi_+^{-1})'+\pi^3\, r^2\, |g_{2,1}|^2 =-\pi^3\, e^{4 C}\, k_1 \,,
\ee
where $(\pi^2/2) e^{4 C}$ is the volume of the resolved $\mathbb{P}^2$ and the near-stack approximation is valid for $r \ll r_0$, where $r_0 = \frac{1}{3}\, e^{2 B+ C}$, as in (\ref{eqn:NS_CYmetric}).

\subsection{Supersymmetry conditions in the near-stack region} \label{subsec:susy-near}

We now apply the supersymmetry conditions for AdS and AdS-like vacua (\S\ref{subsec:SL2Rsusy}) to the near-stack ansatz of \S\ref{subsec:flat}. We assume that the internal spinors $\eta^1$ and $\eta^2$ are singlets under the $SU(3)$ symmetry group, so that $\Omega_2$, $J_2$, $\Theta$ and $\varphi$ are also singlets under the $SU(3)$.

The only invariant one-forms are $\frac{1}{z}\, \mathd z$ and its conjugate, up to a radially-dependent
factor. In order to satisfy the orthonormality conditions $g^{-1}(\Theta,\Theta) = 0$ and $g^{-1}(\Theta,\bar{\Theta}) = 2$, we must take
\be \label{eqn:Theta_ansatz}
  \Theta = e^{- 2 B + i\, \theta}\, \frac{r}{z}\, \mathd z \,,
\ee
for some real $\theta = \theta (r)$. $\Omega_2$ must be a complex two-form that is odd under the involution $z\to -z$, and that satisfies $i_{\Theta}\, \Omega_2 = 0$, $\Omega_2 \wedge \Omega_2 = 0$, with the normalization $\mathrm{vol}_6 = \frac{1}{4} J_1 \wedge \Omega_2 \wedge \bar{\Omega}_2$. Therefore, we must have
\be \label{eqn:Omega_ansatz}
  \Omega_2 = e^{2 C+i\, \xi}\, \frac{z}{r}\, \mathd u^1 \wedge \mathd u^2 \,,
\ee
for some phase factor $\xi = \xi (r)$, and the complex structure defined by $\Omega = \Omega_2 \wedge \Theta$ is integrable by inspection. Then, (\ref{eqn:Theta_ansatz}) and (\ref{eqn:Omega_ansatz}) uniquely determine $J_1$ and $J_2$:
\be \label{eqn:J12_ansatz}
J_1 = \frac{i}{2}\, e^{-4 B} \mathd z \wedge \mathd \bar{z} \;\;,\;\; J_2 = \frac{i}{2}\, e^{2 C} \left(\mathd u^1 \wedge \mathd \bar{u}^1+\mathd u^2 \wedge \mathd \bar{u}^2\right) \,.
\ee
Applying (\ref{eqn:SUSY_ThetaJ}), we see that $J_2$ is closed, since $\varphi= \varphi(r)$ is a singlet scalar, and therefore $c_{\varphi}\, J_1$ is closed by (\ref{eqn:J12_ansatz}). Thus, $C$ is constant and the metric is K\"ahler. By (\ref{eqn:SUSY_dJ}), the solution must be Minkowski ($\mu = 0$).\footnote{This is an artifact of the near-stack approximation: (\ref{eqn:SUSY_ThetaJ}) requires $\mu \ne 0$ for $\varphi \ne 0$ solutions on the $\mathbb{P}^2$ cone. The AdS-like solutions obtained in this section may be viewed as `approximately AdS,' in that they approximate solutions on the $\mathbb{P}^2$ cone with small cosmological constant.}

Applying $\mu = 0$ to the first equation of (\ref{eqn:SUSY_ThetaJ}) and integrating, we find
\be \label{eqn:cos_varphi}
\alpha = c_{\varphi}\, e^{4 A} \,,
\ee
where we fix the $\alpha$ shift symmetry. The middle equation of (\ref{eqn:SUSY_ThetaJ}) gives
\be \label{eqn:Theta_c1}
e^{2 A}\, s_{\varphi}\, \Theta = \frac{c_1}{z}\, \mathd z \,,
\ee
where $c_1$ is a complex constant. Thus,
\be \label{eqn:sin_varphi}
r s_{\varphi}\, e^{2 (A - B) + i \theta} = c_1 \,,
\ee
and $\theta$ must be constant. Combining (\ref{eqn:cos_varphi}, \ref{eqn:sin_varphi}) to eliminate $\varphi$, we find
\be\label{eqn:phipm}
  \Phi_{+}\Phi_{-} = \frac{|c_1|^2}{r^2}\, e^{4(A+B)}\,.
\ee
Henceforth, we assume that $c_1 \neq 0$, since the special case $c_1 = 0$ is just a type B solution, as discussed above. For $c_1 \neq 0$, $\varphi \neq 0, \pi$, and so (\ref{eqn:cos_varphi}) implies that $| \alpha | < e^{4 A}$, and therefore
\be
\Phi_{\pm} > 0\,. \label{eqn:phipm_pos}
\ee

Now consider the decomposition of $G_3$, (\ref{eqn:SUSY_G3ansatz}). $SU(3)$ invariance constrains $\mathcal{G}_1$, $\mathfrak{G}_1$, $\mathcal{G}_{1,1}$, and $\mathfrak{G}_{1,1}$ to vanish. Thus, we decompose:
\be
  G_3 = \mathfrak{g}_{3, 0}\, \Omega_2 \wedge \Theta +\mathfrak{g}_{2, 1}\,
  \Omega_2 \wedge \bar{\Theta} +\mathfrak{g}_{1, 2}\, \bar{\Omega}_2 \wedge
  \Theta +\mathfrak{g}_{0, 3}\, \bar{\Omega}_2 \wedge \bar{\Theta}\,,
\ee
where the $\mathfrak{g}_{p,q}$ are related to the $g_{p,q}$ of \S\ref{subsec:flat} by
\bea
  g_{3, 0} = e^{2 (C- B) + i\, (\xi+\theta)} \mathfrak{g}_{3, 0} & \;\;,\;\; &
  g_{2, 1} = e^{2 (C- B) + i\, (\xi-\theta)} \mathfrak{g}_{2, 1} \,, \nonumber \\
  g_{1, 2} = e^{2 (C- B) - i\, (\xi-\theta)} \mathfrak{g}_{1, 2} & \;\;,\;\; &
  g_{0, 3} = e^{2 (C- B) - i\, (\xi+\theta)} \mathfrak{g}_{0, 3}\,.
\eea
The conditions (\ref{eqn:SUSY_calfrakG10} - \ref{eqn:SUSY_frakG01}) are now trivially satisfied. We use the above decomposition to write out the remaining conditions. Applying (\ref{eqn:cos_varphi}), (\ref{eqn:SUSY_Omega}) becomes
\be \label{eqn:Omega_eqn}
  \mathD_+  \left[ e^{2 A}\, s_{\varphi}\, \Omega_2 \right] = \Phi_+  \left( G_{3, 0} + G_{1, 2} \right)^{\star} - \Phi_-  \left( G_{2, 1} + G_{0, 3} \right)^{\star}\,,
\ee
since $G_3$ is primitive. Next, (\ref{eqn:SUSY_g30_g21}, \ref{eqn:SUSY_g12_g03}) become:
\bea \label{eq:g3-relations}
  \mathfrak{g}_{3, 0} =  -\frac{r\, e^{-4 A}}{4\, c_1}\, \Phi_-\, \frac{i\, \tau'}{\tau_2} & \;\;,\;\; &
  \mathfrak{g}_{2, 1} =  \frac{r\, e^{-4 A}}{4\, \bar{c_1}}\, \Phi_+\, \frac{i\, \tau'}{\tau_2} \,, \label{eqn:g30_g21} \\
  \mathfrak{g}_{1, 2} =  \frac{r\, e^{- 8 A}}{4\, c_1}\, \Phi_+ \Phi_-'  & \;\;,\;\; &
  \mathfrak{g}_{0, 3} = - \frac{r\, e^{-8 A}}{4\, \bar{c}_1}\, \Phi_- \Phi_+' \,, \label{eqn:g12_g03}
\eea
where we have
\be \label{eqn:df_sub}
\mathd f (r) = f' (r)\,\mathd r=f'(r)\, e^{2 A}\, s_{\varphi}\, \mathrm{Re} \left[\frac{r}{c_1}\, \Theta\right] \,,
\ee
so that
\be \label{eqn:partialTheta_sub}
\partial_{\Theta} f = e^{2 A}\, s_{\varphi}\, \frac{r}{2\, c_1}\, f'(r) \,.
\ee

It is convenient to fix an $SL(2, \mathbb{R})$ frame. To do so, we choose some radius $r$ and perform an $SL(2, \mathbb{R})$
transformation to make $\tau'/\tau_2$ imaginary. Thus, at radius $r$, $\mathfrak{g}_{3,0}\, \mathfrak{g}_{0,3}$ and $\mathfrak{g}_{2,1}\, \mathfrak{g}_{1,2}$ are both real. However, in this case the axion equation of motion (\ref{eqn:taueom}) reads $C_0'' = 0$. Thus, in this frame the axion is constant,
so that $i \tau'/\tau_2 = \phi'$. More general solutions can be recovered from this special case by an $SL(2,\mathbb{R})$ transformation.

We evaluate the left-hand side of (\ref{eqn:Omega_eqn}), taking $\mathd\,C_0 = 0$:
\be
\mathD_+ \left[ e^{2 A}\, s_{\varphi}\, \Omega_2 \right] = \mathd \left[ e^{2 A}\, s_{\varphi}\, \Omega_2 \right]
- \frac{1}{2}\, e^{2 A}\, s_{\varphi}\, \mathd \phi \wedge \bar{\Omega}_2 \,.
\ee
Using (\ref{eqn:df_sub}), it is straightforward to check that the second term cancels against the $(1,2)+(0,3)$ component of the right-hand side of  (\ref{eqn:Omega_eqn}).\footnote{This could have been anticipated from the integrability of the complex structure defined by $\Omega_2 \wedge \Theta$.} We are left with
\be
\mathd \left[ e^{2 A}\, s_{\varphi}\, \Omega_2 \right] = \Phi_+\, G_{1, 2}^{\star} - \Phi_-\, G_{0, 3}^{\star} \,. \label{eqn:Omega_eqn1}
\ee
We evaluate the exterior derivative using (\ref{eqn:Omega_ansatz}) and (\ref{eqn:df_sub}):
\be
\mathd \left[ e^{2 A}\, s_{\varphi}\, \Omega_2 \right] =
\frac{1}{2\, c_1}\, e^{2 A - i\, \xi} s_{\varphi} \frac{\mathd}{\mathd r} \Bigl( r e^{2 A+i\, \xi} s_{\varphi} \Bigr)\, \Theta \wedge \Omega_2
+ \frac{r^2}{2\, \bar{c}_1}\, e^{2 A - i\, \xi} s_{\varphi} \frac{\mathd}{\mathd r} \Bigl( \frac{1}{r} e^{2 A+i\, \xi} s_{\varphi} \Bigr)\, \bar{\Theta} \wedge \Omega_2 \,.
\ee
Comparing with (\ref{eqn:g12_g03}), we see that $\xi$ must be a constant. After some manipulation, (\ref{eqn:Omega_eqn1}) becomes
\be \label{eqn:Omega_eqn2}
\frac{1}{r^2}\, \frac{\mathd}{\mathd r} \Bigl( r^2 e^{-4 A}\, \Phi_+ \Phi_- \Bigr) = e^{-8 A}\, \Phi_-^2 \Phi_+'
\;\;,\;\;
r^2\, \frac{\mathd}{\mathd r} \Bigl( \frac{1}{r^2} e^{-4 A}\, \Phi_+ \Phi_- \Bigr) = e^{- 8 A}\, \Phi_+^2 \Phi_-'  \,.
\ee
Defining $\Xi_{\pm} \equiv \Phi_{\pm}^{-1}$, both halves of (\ref{eqn:Omega_eqn2}) reduce to
\be \label{eqn:Xipm}
\frac{2}{r}\, (\Xi_++\Xi_-) = \Xi_-' - \Xi_+'\,.
\ee
Now (\ref{eqn:phipm}) can be rewritten as
\be \label{eqn:B_ansatz}
B = -\frac{1}{4} \log\left[\frac{|c_1|^2}{2\, r^2}\, (\Xi_+ + \Xi_-)\right]\,.
\ee
The supersymmetry conditions for the near-stack ansatz of \S\ref{subsec:flat} are therefore equivalent to (\ref{eqn:phipm_pos}, \ref{eqn:g30_g21}, \ref{eqn:g12_g03}, \ref{eqn:Xipm}, \ref{eqn:B_ansatz}), where, in a general $SL(2,\mathbb{R})$ frame, $\mathD\, e^{i\, \xi} = 0$.\footnote{Like $\Omega_2$, $e^{i \xi}$ carries S-charge $-1/2$.}

\subsection{Solutions in the near-stack limit} \label{subsec:sols-near}

We define
\be \label{eqn:Delta}
 \Delta (r) \equiv \frac{1}{2}  \Biglp  \Xi_- - \Xi_+ \Bigrp \;\;,\;\; f(r) \equiv \Xi_- + \Xi_+=r \Delta' \,,
\ee
where the final equality follows from (\ref{eqn:Xipm}). To further constrain the solution, we must impose the equations of motion. In fact, we will only need two: the $\phi$ equation of motion and the $B$ equation of motion.

First, from the dilaton equation of motion (\ref{eqn:taueom}) we find
\be
  \frac{1}{r}  \frac{\mathd}{\mathd r}  \left[ r \phi' \right] = 8\, e^{4
  A-4 C}  \left( g^{3, 0} g^{0, 3} + g^{2, 1} g^{1, 2} \right) = - \frac{\Xi_+' + \Xi_-'}{\Xi_+ + \Xi_-}\, \phi'  \,,
\ee
in the constant axion frame, so that
\be \label{equ:phiprime}
  \phi' = \frac{2\, c_2}{r\, (\Xi_+ + \Xi_-)}\,,
\ee
where $c_2$ is another constant.

Now consider the $B$
equation of motion (\ref{eqn:Beom}), which  can be written
\be
  \frac{1}{r}  \frac{\mathd}{\mathd r}  \left[ r B' \right]
  =  \frac{\Xi_+' \Xi_-'}{2 \left( \Xi_+ + \Xi_- \right)^2} + \frac{1}{8}\left( \phi' \right)^2\,.
\ee
Inserting (\ref{eqn:B_ansatz}), we find
\be \label{equ:beom}
  \frac{1}{r}  \frac{\mathd}{\mathd r}  \left[ r \left( \Xi_+' + \Xi_-'
  \right) \right] = \frac{\left( \Xi_+' \right)^2 + \left( \Xi_-'
  \right)^2}{\left( \Xi_+ + \Xi_- \right)} - \frac{1}{2}  \left( \Xi_+ + \Xi_-
  \right)  \left( \phi' \right)^2\,.
\ee
It is straightforward to check that the $\Xi_{\pm}$ equations of motion (\ref{eqn:alphaeom}, \ref{eqn:Aeom}) are satisfied provided that (\ref{equ:beom}) and the supersymmetry constraints are obeyed. Substituting (\ref{equ:phiprime}) and (\ref{eqn:Delta}), (\ref{equ:beom}) takes the form
\be\label{equ:feq}
  \frac{1}{r}  \frac{\mathd}{\mathd r}  \left[ r f' \right] = \frac{1}{f}
  \left[ \frac{1}{2}  \left( f' \right)^2 + \frac{2}{r^2}\, (f^2 - c_2^2)\right]\,.
\ee
A general solution is of the form\footnote{To see this, first differentiate (\ref{equ:feq}) with respect to $r$, obtaining a third-order equation whose nonlinear terms can be eliminated using  (\ref{equ:feq}). The result gives $\frac{1}{r}  \frac{\mathd}{\mathd r}  \left[ r g' \right]  =  \frac{4}{r^2} g$ where $g (r) \equiv r f' (r)$. The solution is then easy to guess.}
\be
  f (r) = c_3 r^2 + c_4 r^{- 2} + c_5\,.
\ee
Substituting this result into (\ref{equ:feq}), one obtains
\be \label{eqn:c2_cons}
  c_2^2 = c_5^2 - 4\, c_3\, c_4\,.
\ee
We integrate once more to find
\be
  \Delta (r) = \frac{1}{2}\, c_3\, r^2 - \frac{1}{2}\, c_4\, r^{- 2} + c_5\,
  \log\, r + c_6\,.
\ee
Thus,
\be \label{eqn:Xi_sol}
\Xi_+ = \left(c_5/2-c_6\right)+c_4\, r^{-2}-c_5\,\log\,r \;\;,\;\; \Xi_- = \left(c_5/2+c_6\right)+c_3\, r^2+c_5\,\log\,r \,,
\ee
and (\ref{equ:phiprime}) becomes:
\be
  \phi' = \frac{2\, c_2}{r \left( c_3\, r^2 +
  c_4\, r^{- 2} + c_5 \right)}\,.
\ee
This can be integrated to give
\be\label{eq:dilaton}
  \phi (r) = \phi_0 + \log \left[ \frac{\left(c_5 + c_2 \right) r^2 + 2\, c_4}{\left( c_5 -c_2 \right) r^2 + 2\, c_4} \right]\,,
\ee
where we use (\ref{eqn:c2_cons}).

The full solution is given by (\ref{eqn:Xi_sol}, \ref{eq:dilaton}, \ref{eqn:g30_g21}, \ref{eqn:g12_g03}, \ref{eqn:B_ansatz}), where $\Xi_{\pm} \equiv \Phi_{\pm}^{-1}$ as above. It is straightforward to check that the remaining equations of motion (\ref{eqn:g30eom} - \ref{eqn:g03eom}, \ref{eqn:BCcons}) are automatically satisfied. An interesting special case is where $c_2 = 0$, or $c_5^2 = 4\, c_3\, c_4$, so that $\phi' = 0$. This implies that $\mathfrak{g}^{3, 0} = \mathfrak{g}^{2, 1} = 0$, but $\mathfrak{g}^{1, 2}$ and $\mathfrak{g}^{0, 3}$ are nonvanishing. These are explicit examples of constant $\tau$ solutions that are not type B solutions.

\subsection{Beyond the near-stack limit} \label{subsec:P2_susy}

We now briefly consider the extension of our methods to the full $\mathbb{P}^2$ cone. We must have
\be
  \Theta = \frac{1}{r}\, e^{- 2\, B + i\, \theta}\, \partial\, r^2 \,,
\ee
which agrees with (\ref{eqn:Theta_ansatz}), since $r^2 \to |z|^2$ in the near-stack limit. We compute
\bea
\mathd \left[e^{2 A}\,s_{\varphi}\,\Theta\right]
& = & \frac{i}{r} \frac{\mathd}{\mathd r} \left[r\, s_{\varphi}\, e^{2 (A-B)+i\,\theta}\right] \chi_{1,1}
+6\, i\, r\, s_{\varphi}\, e^{2 (A-B)+i\,\theta}\,\omega_{1,1} \nonumber \\
& = & 2\, i\, \mu \left(c_{\varphi}\, e^{-4 B}\, \chi_{1,1}+e^{2 C}\, \omega_{1,1}\right)\,, \label{eqn:P2theta_cond}
\eea
where we used (\ref{eqn:omegachi_def}, \ref{eqn:SUSY_ThetaJ}) and the decomposition $J_1 =  \frac{i}{2}\, \Theta\wedge\bar{\Theta} = e^{-4 B} \chi_{1,1}$ and therefore $J_2 = J - J_1 = e^{2 C}\, \omega_{1,1}$. Comparing the terms proportional to $\omega_{1,1}$ in (\ref{eqn:P2theta_cond}), we find
\be \label{eqn:P2mueqn}
3\, r\, s_{\varphi}\, e^{2 (A-B)+i\,\theta}=\mu\, e^{2 C}\,.
\ee
Thus, for $s_{\varphi} \ne 0$, we must take $\mu\ne0$, and the solution is always AdS. Using methods similar to those in \S\ref{subsec:susy-near}, one can check that supersymmetric AdS solutions with global $SU(3)$ symmetry of this type do exist. The resulting ODEs are nonlinear and difficult to solve except in certain special cases, and we defer further consideration of this problem to a later work.

Comparing (\ref{eqn:P2mueqn}) with (\ref{eqn:sin_varphi}), we see that $c_1$ of the near-stack ansatz is related to $\mu$ via
\be \label{eqn:c1mu_rel}
c_1=\frac{\mu}{3}\,e^{2\, C_{\rm ns}}\,,
\ee
where $C_{\rm ns}$ is the constant near-stack value of $C$.

\section{Geometry of the Near-stack Solution}\label{sec:solprops}

\subsection{Singularity structure}

The solution found in \S\ref{sec:solution} depends on several parameters: the $c_i$ ($i = 1 \ldots
6$), $g_s = e^{\phi_0}$, $C$,
$\xi$,
and $C_0$.
The $c_i$ must obey the relation (\ref{eqn:c2_cons}), and
in particular we must have
\be \label{eqn:c5_cons}
  c_5^2 \geqslant 4 c_3 c_4 \,.
\ee
The explicit solution is given by
\bea
  \Xi_+ (r) = (c_5/2 - c_6) + c_4\, r^{-2} - c_5 \log r & \;\;,\;\; &
  \Xi_- (r) = (c_5/2+ c_6) + c_3 r^2 + c_5 \log r \,,\\
  \tau = C_0 + \frac{i}{g_s}  \left( \frac{(c_5 - c_2) r^2 + 2 c_4}{(c_5 +
  c_2) r^2 + 2 c_4} \right) & \;\;,\;\; &
  B = \frac{1}{4} \log \left[ \frac{2\, r^4}{|c_1|^2  \left(c_4 + c_5 r^2+c_3 r^4\right)} \right] \,,
\eea with fluxes specified in (\ref{eqn:g30_g21}, \ref{eqn:g12_g03}). The spinor angle $\varphi$ varies with radius,
\be
\cos\,\varphi = \alpha\, e^{-4 A} = \frac{-c_4 + 2\, c_6\, r^2 + 2\, c_5\, r^2 \log r+ c_3\, r^4}{c_4 + c_5\, r^2 + c_3\, r^4}\,,
\ee
so the solution has dynamic $SU(2)$ structure.

For the $\varphi \neq 0, \pi$ (i.e.\ non-type B) case that we are considering, the $c_i$ must all be finite or vanishing, and the solution is regular and supersymmetric whenever the $\Xi_{\pm}$ are both positive. We now show that a singularity (i.e.\ a zero in either $\Xi_+$ or $\Xi_-$)
always occurs at finite radius. Consider the sum (cf. (\ref{eqn:Delta})):
\be
r^2 f(r) = r^2 (\Xi_+ + \Xi_-) = c_4 + r^2\, c_5 + r^4\, c_3 \,.
\ee
This must be positive as a necessary but insufficient condition for regularity and supersymmetry.

Assume that there exists a solution that is smooth at all finite radii. To maintain regularity in the large and small $r$ regions, we must have $c_4 \geqslant 0$ and $c_3 \geqslant 0$, and moreover, since the discriminant of the quadratic polynomial $r^2 f(r)$ is nonnegative by (\ref{eqn:c5_cons}), $c_5$ must be nonnegative, or else $\Xi_+ + \Xi_-$ acquires a root at some finite radius. However, under these conditions $\Xi_+$ will become negative at small $r$ unless $c_5 = 0$. If this is the case, then either $c_3$ or $c_4$ must vanish by (\ref{eqn:c5_cons}), and $c_6$ must be respectively either positive or negative to obtain a regular solution anywhere. In either case, one of $\Xi_+$ or $\Xi_-$ is constant, whereas the other crosses zero at some finite radius. Thus, a singularity will always occur at some radius. The constraint (\ref{eqn:c5_cons}) plays a crucial role in this argument.


A singularity at finite radius is always of the
$e^{A} \to \infty$ type. It is straightforward to check that curvature invariants diverge and the singularity is physical. Moreover, horizons, characterized by $e^{A} \to 0$, can only occur for $r \to 0$ and/or for $r \to \infty$, so the singularity is naked.

We now classify the available regions of parameter space for which a regular solution exists at some radius. Clearly we must have $f = \Xi_+ + \Xi_- > 0$. However, this is also sufficient at any given point for some choice of $c_6$, since we can always set $\Xi_+ = \Xi_-$ at any point of interest by adjusting $c_6$. If either $c_3$ or $c_4$ is positive, then $f$ is positive at large or small $r$, respectively, and $c_6$ can be chosen such that a regular region exists. For solutions with neither $c_3$ nor $c_4$ positive, one can check that a region of positive $f$ exists so long as $c_5$ is positive and the inequality $c_5^2 \geqslant 4 c_3 c_4$ is not saturated.

\begin{figure}[h!]
\centering
\includegraphics[width=4in]{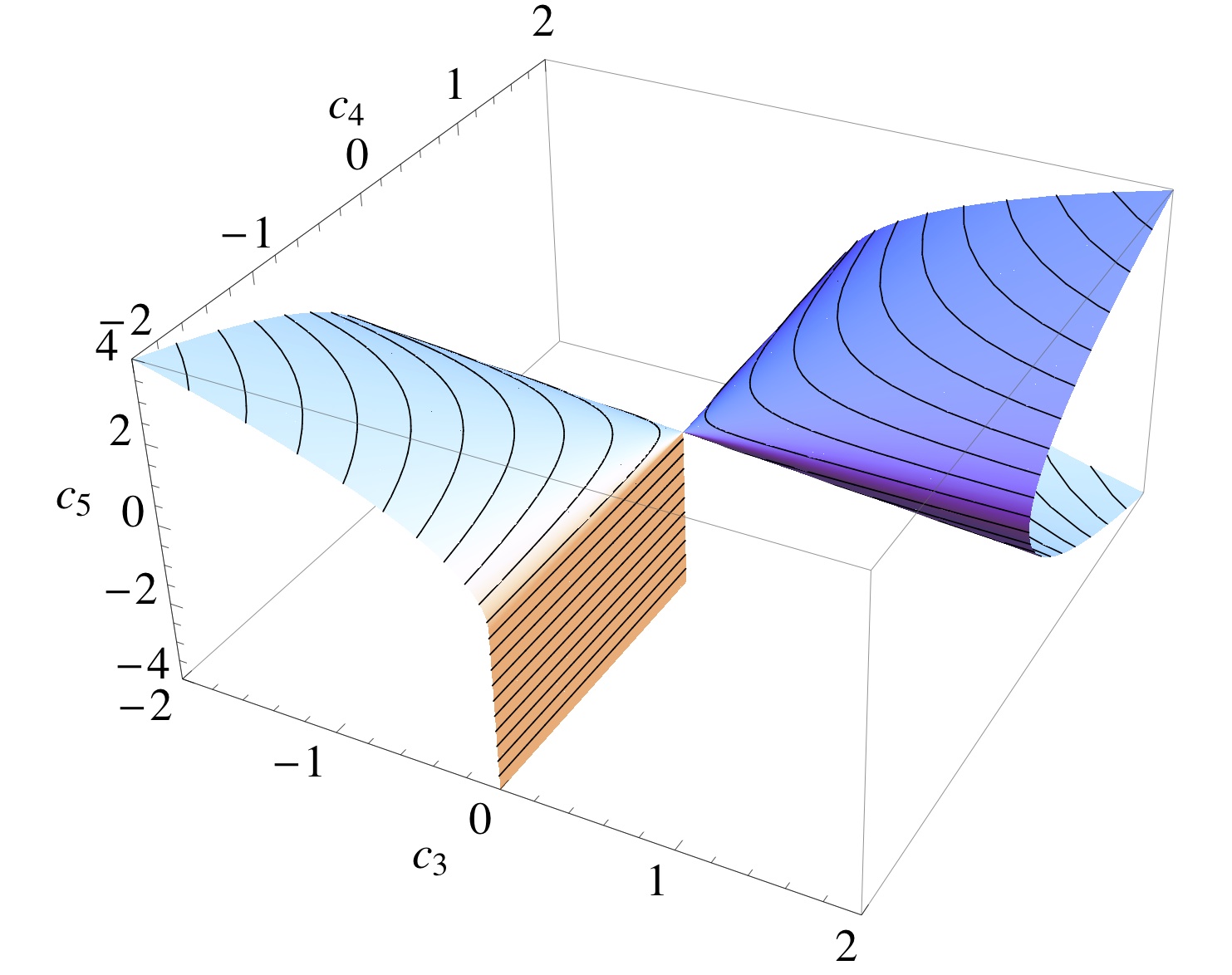}
\caption{$(c_3,c_4,c_5)$ parameter space. The region inside the cones is excluded. The surface of the negative cone is also excluded, whereas the surface of the positive cone consists of constant dilaton solutions. Hyperbolae of constant $c_3\,c_4$ are related by radial rescaling.}
\label{fig:general-space}
\end{figure}

The space of available $(c_3, c_4, c_5)$ can be imagined as $\mathbb{R}^3$ minus two cones,
a `positive' cone in the region $c_3, c_4 > 0$ given by $c_5^2 < 4 c_3 c_4$, and a `negative' cone in the region $c_3, c_4 < 0$ given by $c_5 \leqslant \sqrt{4 c_3 c_4}$. The surface of the positive cone is available, and consists of the constant dilaton solutions ($c_2=0$), whereas the surface of the negative cone is unavailable. Not all points in this space are physically distinct, as radial rescalings $r \rightarrow \lambda\, r$ trace out hyperbolae $c_3\, c_4 = \tmop{const}$. The $(c_3,c_4,c_5)$ parameter space is depicted in Figure \ref{fig:general-space}.

\subsection{Constant dilaton solutions}\label{subsec:tau-const}

The space of solutions is large and complicated, and without an explicit mechanism for resolving the singularity and lacking an asymptotic AdS region for comparisons with a boundary field theory, it is hard to anticipate which of these solutions will be realized physically.
We will focus on one of the simplest classes, in which the dilaton is constant $(c_2=0)$.\footnote{One motivation for considering this class of solutions is that string loop corrections can be controlled parametrically by taking $g_s \ll 1$. Although $g_s$ is still available as a control parameter, the situation is more complicated when the dilaton runs.} These are the solutions
that lie on the positive
cone $c_5^2 = 4 c_3 c_4$ with $c_3, c_4 \geqslant 0$, as noted above. We will not consider the special cases where either $c_3$ or $c_4$ vanishes, so $c_5 \neq 0$ in general. It is convenient to reparameterize:
\be
  c_3 = \frac{|c_5|}{2\, r_{\star}^2} \;\;,\;\;
  c_4 = \frac{|c_5|\, r_{\star}^2}{2} \;\;,\;\;
  c_6 = \frac{1}{2}\, \delta\, |c_5| - c_5 \log r_{\star} \,,
\ee
where $c_5$ is related to the D3-brane charge.
Using (\ref{eqn:P2_QD3Page}), we compute
\be\label{eq:Q-c5}
Q = \pi^3\, e^{4 C} c_5\,,
\ee
where $Q$ is the Page charge.
We find that
\be
  \Xi_+ (r)  =   \frac{|Q|}{2\, \pi^3}\,e^{-4\, C} \left[ \frac{r_{\star}^2}{r^2} +\mathrm{sgn}(Q)\, (1 - 2 \log
  r / r_{\star}) - \delta \right] \,, \label{eqn:Xi_constdil}
\ee
\be
  \Xi_- (r)  =  \frac{|Q|}{2\, \pi^3}\,e^{-4\, C} \left[ \frac{r^2}{r_{\star}^2} +\mathrm{sgn}(Q)\, (1 + 2 \log
  r / r_{\star}) + \delta \right] \,, \label{eqn:Xi_constdil2}
\ee
where $\mathrm{sgn}(Q)$ is the sign of $Q$.

We consider the case of positive and negative $Q$ separately. For positive $Q$, $\Xi_-$ is a strictly increasing function of $r$, whereas $\Xi_+$ is a
strictly decreasing function of $r$.
Then, $f = \Xi_+ + \Xi_- \propto \frac{(r^2 + r_{\star}^2)^2}{r^2 r_{\star}^2}$ is positive everywhere, so the solution can be made regular in any region by an appropriate choice of $\delta$. However, $\Xi_+ \rightarrow - \infty$ as $r \rightarrow \infty$ and $\Xi_- \rightarrow - \infty$ as $r \rightarrow 0$, so the solution is only valid between two radii $r_1$ and $r_2 > r_1$ where $\Xi_-$ and $\Xi_+$ cross zero, respectively. One can easily show that
\be
  r_2 = r_{\star}  \left[ W_0 (e^{\delta - 1}) \right]^{- 1 / 2} \;\;,\;\;
  r_1 = r_{\star}  \left[ W_0 (e^{- \delta - 1}) \right]^{1 / 2} \,,
\ee
where $W_0$ is the main branch of the Lambert W-function. In particular, the ratio of the two
scales is
\be
  (r_2 / r_1)^2 = \left[W_0 (e^{\delta - 1}) W_0 (e^{- \delta - 1})\right]^{-1}\,.
\ee
This ratio is minimized at $\delta = 0$, where it takes the value $r_2 / r_1 =
\left[ W_0 (1 / e) \right]^{-1} \simeq 3.59$. For $| \delta | > 0$, the ratio
increases, and asymptotically for large $| \delta |$, we find
\be
  (r_2 / r_1)^2 \rightarrow | \delta |^{- 1} e^{1 + | \delta |}\,,
\ee
so the region of regularity can be made very large for modest values of $\delta$.

For the case of negative $Q$, $\Xi_+$ and $\Xi_-$ have a single minimum at $r_{\star}$, and $f
\propto \frac{(r^2 - r_{\star}^2)^2}{r^2 r_{\star}^2} = 0$ at $r = r_{\star}$, so the solution can be made regular anywhere but at $r_{\star}$. For any choice of $\delta$, the solution is regular for $r > r_2$ and $r < r_1$, where for $\delta > 0$ both singularities are due to $\Xi_+$ crossing zero, and for $\delta < 0$ both are due to $\Xi_-$ crossing zero. The radii of the singularities for $\delta > 0$ are
\be
  r_1 = r_{\star}  \left[ - W_{- 1} (- e^{- 1 - \delta}) \right]^{- 1 /2} \;\;,\;\;
  r_2 = r_{\star}  \left[ - W_0 (- e^{- 1 - \delta}) \right]^{- 1 / 2} \,,
\ee
where $W_{-1}$ is the lower branch of the Lambert W-function. For $\delta < 0$, we have instead
\be
  r_1 = r_{\star}  \left[ - W_0 (- e^{\delta - 1}) \right]^{1 / 2} \;\;,\;\;
  r_2 = r_{\star}  \left[ - W_{- 1} (- e^{\delta - 1}) \right]^{1 / 2} \,.
\ee

For the special case $\delta = 0$, $r_1 = r_2 = r_{\star}$, and the solution is regular everywhere else. This special case has interesting properties. For instance, the spinor angle is finite everywhere:
\be
\cos \varphi = \frac{r^4-r_{\star}^4-4\, r^2 \log (r/r_{\star})}{(r^2-r_{\star}^2)^2}\,.
\ee
We see that $\varphi$ interpolates between $\varphi = 0$ as $r \to \infty$ and $\varphi = \pi$ as $r \to 0$, passing through a type-changing locus ($\varphi = \pi/2$)
coincident with the singularity at $r=r_{\star}$.

\begin{figure}[tbp]
\centering
\includegraphics[width=2.5in]{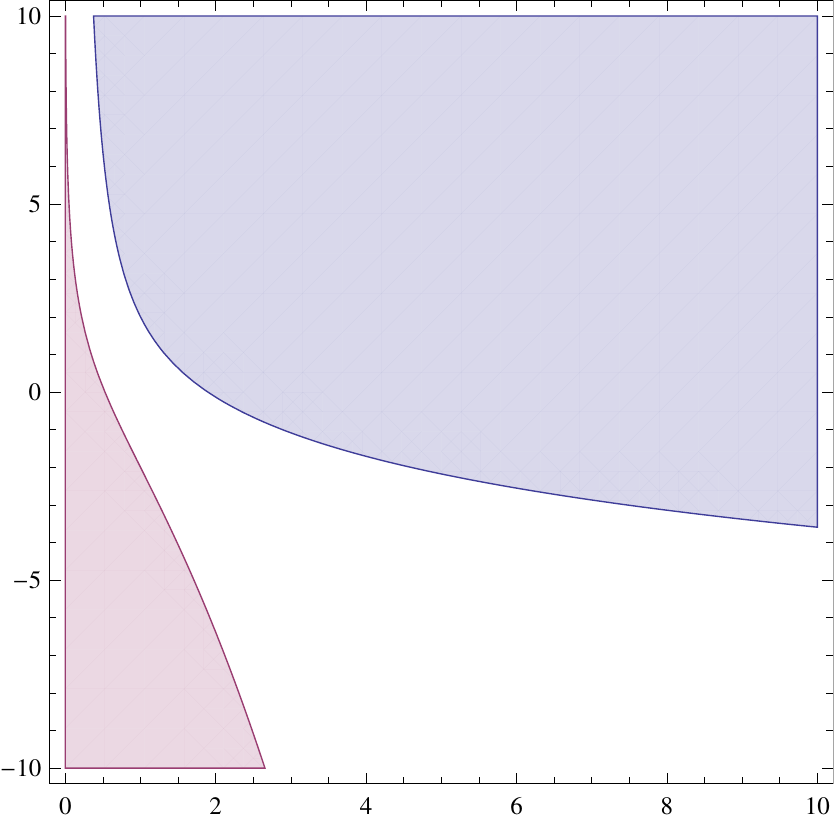}
\hspace{15mm}
\includegraphics[width=2.5in]{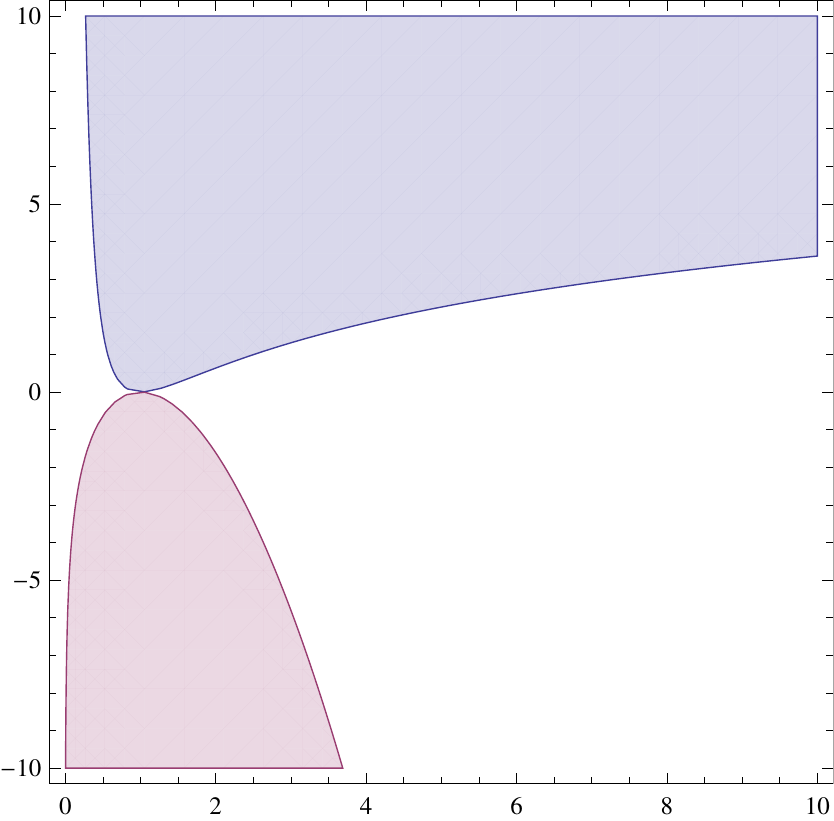}
\caption{The singularity structure of the constant dilaton solutions. The horizontal axis is $r/r_{\star}$ and the vertical axis is
$\delta$. The nonsingular region is unshaded and $\Xi_+$ ($\Xi_-$) is negative in the
blue (red) region. The first (second) plot
corresponds to positive (negative) $Q$.}
\label{fig:const-tau}
\end{figure}

The singularity structure of constant dilaton solutions is illustrated in Figure \ref{fig:const-tau} for both positive and negative $Q$. The special case discussed above corresponds to $\delta=0$ in the second plot of the figure.

\subsection{Domain of validity of the near-stack solutions}\label{subsec:validity}

While we have obtained a family of closed-form solutions to the supergravity equations, our result is subject to corrections from a number of sources.
Though string loop corrections can be controlled parametrically for constant dilaton solutions, the solutions we have obtained are always singular at some finite radius, and so $\alpha'$ corrections will invariably be large in some regions. Furthermore, the near-stack approximation employed in \S\ref{subsec:flat} also has a finite region of validity. Generically we expect an interplay between these two sources of corrections, as curvatures should fall off at large distances whereas the near-stack approximation is valid at short distances. We now estimate the size of corrections from various sources, and show that in some regions of parameter space we can parametrically suppress all corrections in some finite interval $r_{\rm near} < r < r_{\rm far}$.

\subsubsection{Corrections to the near-stack limit}

Corrections to the near-stack limit of \S\ref{subsec:flat} come in two forms. At large distances from the stack wrapping the resolved $\mathbb{P}^2$, the resolution appears as a small perturbation to the geometry, and the seven-branes appear to be
codimension six sources, rather than codimension two sources as they do in the near-stack limit. This effect manifests itself in the Einstein equations for the $\mathbb{P}^2$ cone as extra terms of order $(r / r_0)^2$ and $(r / r_0)^4$, where $r_0$ is the radial scale of the blow-up, as in (\ref{eqn:NS_CYmetric}); that is, one can show that the Einstein equations (\ref{eqn:Ceom}, \ref{eqn:Beom}, \ref{eqn:BCcons}), written in the form
\be
r^2\, B''=\ldots \;\;,\;\; r^2\, C''=\ldots \;\;,\;\; 2\, r\, C'=\ldots\,,
\ee
receive corrections of the form $r^2\, e^{- 4 B - 2 C}$ and $r^4\, e^{- 8 B - 4 C}$.  (In the remainder of this section, we will omit numerical factors in expressions of the form $a\ll b$.)

Assuming that dimensionless terms of the form $r B'$, etc., are of order unity, the extra terms are suppressed if
\be \label{eqn:NSprebound1}
   r^2\, e^{- 4 B - 2 C} \ll 1\,.
\ee
By (\ref{eqn:NS_CYmetric}), this is equivalent to the requirement $(r / r_0)^2 \ll 1$.

The other source of corrections to the near-stack limit is the finite cosmological constant $\Lambda = - 3\, |\mu|^2$. This sources new terms in the
Einstein equations and the warp factor equation of the form $r^2\, e^{- 4 (A + B)}\, \Lambda$, as in (\ref{eqn:Aeom}, \ref{eqn:Ceom}, \ref{eqn:Beom}, \ref{eqn:BCcons}). Applying the relation (\ref{eqn:c1mu_rel}), we obtain the requirement
\be \label{eqn:NSprebound2}
 r^2\, e^{- 4 (A + B + C)}\, |c_1|^2 \ll 1 \,.
\ee
Using (\ref{eqn:phipm}, \ref{eqn:B_ansatz}) the conditions (\ref{eqn:NSprebound1}, \ref{eqn:NSprebound2}) can be rewritten in the form:
\be \label{eqn:NSbound1}
 |c_1|^2\, e^{-2 C} \left(\Xi_+ + \Xi_- \right) \ll 1 \;\;,\;\;  |c_1|^4\, e^{-4 C}\, \Xi_+\, \Xi_- \ll 1\,.
\ee
However, for $\Xi_{\pm}>0$,
\be
|\Xi_+\, \Xi_-| \le  \frac{1}{4}\, (\Xi_+ + \Xi_-)^2\,.
\ee
Therefore, for supersymmetric solutions, the second bound in (\ref{eqn:NSbound1}) is implied by the first.

For solutions with nonvanishing $Q$, it is convenient to extract an overall scale from $\Xi_{\pm}$:
\be \label{eqn:Fpm}
\Xi_{\pm} = \frac{|Q|}{2\, \pi^3}\, e^{-4\, C}\, F_{\pm} \,,
\ee
where for constant dilaton solutions, the $F_{\pm}$ are given by the bracketed terms in (\ref{eqn:Xi_constdil}, \ref{eqn:Xi_constdil2}). The bound (\ref{eqn:NSbound1}) becomes
\be \label{eqn:NSbound2}
\zeta\, |Q| \, (F_+ + F_-) \ll 1\,,
\ee
where $\zeta \equiv  |c_1|^2\, e^{-6\, C}$. The constants $c_1$ and $e^{C}$ naturally appear in physical quantities in this combination, since both can be rescaled by redefining the four-dimensional coordinates $x^{\mu}\to\lambda\, x^{\mu}$ and absorbing the rescaling into the warp factor and the six-dimensional metric, leaving the ten-dimensional metric invariant; $\zeta$ is invariant under this rescaling.

\subsubsection{Curvature corrections}

To estimate the size of $\alpha'$ corrections, we compute the ten-dimensional Riemann tensor components in Einstein frame. We assume that corrections can be suppressed if these components (expressed in an orthonormal basis) can be made parametrically small in units of $\alpha'$. The size of $\alpha'$ corrections in Einstein frame will actually depend on $g_s$, since the F-string tension depends on $g_s$ in this frame. However, if we restrict our attention to constant dilaton solutions, and fix some small value of $g_s$ to suppress string loop corrections, this will only modify the curvature scale at which corrections set in by a fixed factor, and parametric control of the Einstein-frame curvature is still sufficient to suppress corrections. We will not consider corrections involving the other supergravity fields.\footnote{One might hope that these corrections are also suppressed when the Riemann components are small, but not all of these corrections are known, even at leading order, so more detailed estimates may be misguided.}

The ten-dimensional Riemann tensor for the metric (\ref{eqn:10dmetric}) can be computed using standard methods. Expressing everything in an orthonormal basis, one finds terms of the form $e^{-2 A}\, [R_{(4)}]_{\mu \nu \rho \sigma}$, $e^{2 A}\, [R_{(6)}]_{m n p q}$, $e^{2 A}\, (\nabla_m A)(\nabla_n A)$, and $e^{2 A}\, \nabla_m \nabla_n A$, where $[R_{(4)}]_{\mu \nu \rho \sigma}$ are the components of the four-dimensional Riemann tensor computed from $h_{\mu \nu}$, $[R_{(6)}]_{m n p q}$ are the components of the (unwarped) six-dimensional Riemann tensor computed from $g_{m n}$, and $\nabla$ is the connection computed from $g_{m n}$. We assume that the near-stack approximation holds in the region of interest, and use the ansatz of \S\ref{subsec:flat} to compute
\bea
e^{2 A} [R_{(6)}]_{\hat{r} \hat{\psi} \hat{r} \hat{\psi}} = 2\, e^{2 A + 4 B} \left(B''+\frac{1}{r} B'\right) & \;\;,\;\; &
e^{2 A}\, (\nabla_{\hat{r}} A)(\nabla_{\hat{r}} A) = e^{2 A+4 B} (A')^2 \,, \nonumber \\
e^{2 A}\, \nabla_{\hat{r}} \nabla_{\hat{r}} A = e^{2 A+4 B} (A''+2 A' B') & \;\;,\;\; &
e^{2 A}\, \nabla_{\hat{\psi}} \nabla_{\hat{\psi}} A = e^{2 A+4 B}\left(\frac{1}{r} A'-2 A' B'\right)\,, ~~~~ \label{eqn:internal_Riemann}
\eea
where all other terms vanish. If we repeat this computation for the full $\mathbb{P}^2$ geometry, we obtain extra terms which
are suppressed in the near-stack limit, but which scale similarly to those above. However, we also obtain a contribution from the cosmological constant which scales distinctly:
\be  \label{eqn:external_Riemann}
e^{-2 A} [R_{(4)}]^{\mu}_{\;\;\nu \rho \sigma} =-|\mu|^2\, e^{-2 A} \left(\delta^{\mu}_{\rho}\, \eta_{\nu \sigma}-\delta^{\mu}_{\sigma}\, \eta_{\nu \rho} \right) \,.
\ee

Taking $r A'$ and other dimensionless derivatives of $A$ and $B$ to be
of order unity, all the components in (\ref{eqn:internal_Riemann}) are of the same order, as determined by the prefactor, whose square is:
\be \label{eqn:curv_prefactor}
\frac{1}{r^4}\, e^{4 A + 8 B} = \left(\frac{|c_1|^4}{2}\, \Xi_+\, \Xi_-\, (\Xi_+ + \Xi_-)\right)^{-1} =\left(\frac{|Q|}{2 \pi^3}\right)^{-3}\, \zeta^{-2}\, \frac{2}{F_+\, F_-\, (F_+ + F_-)} \,,
\ee
where we have
used (\ref{eqn:Fpm}). By contrast, the square of the curvature induced by the cosmological constant, (\ref{eqn:external_Riemann}), is:
\be \label{eqn:cc_curv}
|\mu|^4\, e^{-4 A} = \frac{81\, \zeta^2\, |Q|}{2 \pi^3}\, \frac{2\, F_+\, F_-}{F_+ + F_-} \,,
\ee
using (\ref{eqn:c1mu_rel}).

\subsubsection{Regions of parametric control}

To suppress $\alpha'$ corrections, we should parametrically suppress (\ref{eqn:curv_prefactor}, \ref{eqn:cc_curv}) while also satisfying (\ref{eqn:NSbound2}).
Rewriting $\zeta |Q| = \epsilon_1$ and $|Q|^{-3} \zeta^{-2} = \epsilon_2$, we see that near-stack corrections can be controlled by taking $\epsilon_1 \ll 1$,
while curvature corrections can be controlled by taking $\epsilon_2$ to be well below some fiducial curvature scale. The curvature (\ref{eqn:cc_curv}) is then suppressed by $\epsilon_1^4\, \epsilon_2 $. However, in this limit $|Q| = 1/(\epsilon_1^2 \epsilon_2)$, so the D3-brane charge must be taken to be large.

In a physical solution, we expect that the D3-brane charge is determined by the worldvolume dynamics on the seven-brane stack, and is not a free parameter. If we add D3-branes near the tip, the 3-7 strings become light and introduce
light matter into the worldvolume theory, precluding gaugino condensation.
Thus, $|Q|$ is not a free parameter, and we must look elsewhere for parametric control of the curvature. Comparing (\ref{eqn:NSbound2}) and (\ref{eqn:curv_prefactor}), we see that control may be possible for $F_{\pm} \gg 1$ with $\zeta$ scaled appropriately (e.g.\ $\zeta \propto F^{-4/3}$). For constant dilaton solutions, (\ref{eqn:Xi_constdil}) gives
\be
F_+ + F_- = \frac{r_{\star}^2}{r^2}+\frac{r^2}{r_{\star}^2} +2\, \mathrm{sgn}(Q)\,.
\ee
Thus, $F_{\pm} \gg 1$ requires $r \gg r_{\star}$ or $r \ll r_{\star}$.

Solutions in the region $r \ll r_{\star}$ are unlikely to be
physical for the following reason. Referring to Figure~\ref{fig:const-tau}, we see that such a solution can only be regular for $r < r_s$ where $r_s \le r_{\star}$. The large-distance singularity at $r = r_s$ is not obviously due to localized sources, and should be removed by compactification, rather than by curvature corrections. However, corrections to the near-stack limit (the first step towards compactification) \emph{decrease} with increasing $r$ in the region $r < r_{\star}$, since $\Xi_+ + \Xi_-$ has a single minimum at $r = r_{\star}$.

Thus, we restrict our attention to the $r \gg r_{\star}$ case. For simplicity, we consider the case where $\Xi_+$ and $\Xi_-$ cross at some large radius $r_{\rm eq} \gg r_{\star}$. Thus, at $r_{\rm eq}$, $F_+ = F_- \approx  r_{\rm eq}^2/(2\, r_{\star}^2)$. In this limit, one can show that the curvature terms are indeed suppressed. At leading order in $r_{\star} \ll r_{\rm eq}, r$, we find
\be
r^2 A''(r) \to \frac{6\, r^2\, r_{\rm eq}^2 - r_{\rm eq}^4}{2\, (2\, r^2 - r_{\rm eq}^2)^2} \;\;,\;\; r A'(r) \to \frac{- r_{\rm eq}^2}{2\, (2\, r^2 - r_{\rm eq}^2)} \;\;,\;\; r^2 B''(r) \to -\frac{3\, r_{\star}^2}{r^2} \;\;,\;\; r B'(r) \to \frac{r_{\star}^2}{r^2} \,,
\ee
so the derivatives are of order one or smaller for $r \gtrsim r_{\rm eq}$, and the size of the Riemann components is determined by the prefactor (\ref{eqn:curv_prefactor}). Thus, our previous arguments apply, and we conclude that both curvature and near-stack corrections can be suppressed at $r \sim r_{\rm eq}$.\footnote{For $r_{eq} \gg r_{\star}$ the singularity occurs at $r_{eq}/\sqrt{2}$, but the ten-dimensional distance between $r_{\rm eq}$ and the singularity is proportional to $r e^{-2 B-A}$, and will be large in string units when the curvatures are small.} Depending on the parameters, near-stack corrections will become important at some $r_{\rm far} > r_{\rm eq}$ and curvature corrections at some $r_{\rm near} < r_{\rm eq}$. Near $r = r_{\rm eq}$ both types of
corrections are small, but $\varphi \sim \pi/2$ and the solution cannot be described by perturbations about a type B background.

As illustrated by this example, the solutions we have obtained
describe physics inaccessible to previous approaches. We stress that the above is not a complete classification of regions of parametric control. We leave further exploration of the large parameter space of solutions, including solutions where the dilaton runs, to a future work.

\section{Towards the Physics of the Solutions}\label{sec:future}

Our choice to study four D7-branes atop an O7-plane on a rigid four-cycle was strongly motivated by the fact that the corresponding seven-brane gauge theory confines
in the infrared, but our analysis so far has exclusively involved ten-dimensional supergravity, without any input of gauge theory physics.
We have obtained a family of exact noncompact solutions parameterized by
a number of
integration constants,
but we expect that some of these constants
are actually {\it{fixed}} by a proper inclusion of nonperturbative source terms localized on the seven-branes  (cf. \cite{Koerber:2007xk},\cite{BDKKM}),
or by matching to the
seven-brane gauge theory.

In this section we present a preliminary analysis of the relation between the solutions obtained
in \S\ref{sec:solution} and the dynamics of the four-dimensional gauge theory on seven-branes in the $\mathbb P^2$ cone. Potential applications of our results to the ten-dimensional description of KKLT vacua and geometric transitions for seven-branes are also discussed.

\subsection{Gaugino condensation in supergravity}\label{subsec:various}

Let us start by briefly discussing the gauge theory supported on the seven-brane stack wrapping
the rigid holomorphic $\mathbb P^2$. An important subtlety in obtaining the worldvolume gauge theory is the following. Since $\mathbb P^2$ is not spin, wrapping D7-branes on it gives rise to global anomaly~\cite{Minasian:1997mm,Freed:1999vc}, whose cancellation requires a nontrivial gauge bundle that will break $SO(8)$ down to (at most) $U(4)$.
While it is not clear to us how the anomaly constraint is modified in the presence of the O7-plane, if we assume that the anomaly does persist,
the resulting gauge group need not be asymptotically free.

In this work we will assume that the seven-brane gauge group
generates a gaugino condensate in the infrared, postponing a proper treatment of the Freed-Witten anomaly
to future work.  Fortunately, the methods we
have developed apply equally well to the Calabi-Yau cone over
$\mathbb{P}^{1}\times \mathbb{P}^{1}$.  The gauge theory analysis
there is simpler, because $\mathbb{P}^{1}\times
\mathbb{P}^{1}$ is spin, but the supergravity analysis becomes
slightly more involved than in the $\mathbb{P}^2$ cone.

Let us now turn to gaugino condensation. Denoting the dynamical scale by $\Lambda$ and the dual Coxeter number of the nonabelian group by $\mathfrak{c}_2$ (not to be confused with the integration constant $c_2$), we have
\be\label{eq:vev-gaugino}
\langle \lambda \lambda \rangle \sim \alpha_{\mathfrak{c}_2} \Lambda^3\,,
\ee
where $\alpha_{\mathfrak{c}_2}$ is a $\mathfrak{c}_2$-th root of unity. The nonperturbative superpotential is also proportional to the gaugino bilinear $\langle \lambda \lambda \rangle$. Recall that the $U(1)_R$ symmetry that acts on the gauginos is anomalous at the quantum level:
\be
\lambda \to e^{i \theta} \lambda\;\;,\;\;\tau_{YM} \to \tau_{YM} +  \frac{\mathfrak{c}_2}{\pi}\, \theta\,.
\ee
The exact symmetry is thus reduced to a discrete $\mathbb Z_{2 \mathfrak{c}_2}$. Moreover, this is spontaneously broken to $\mathbb Z_2$ by (\ref{eq:vev-gaugino}), leading to $\mathfrak{c}_2$ inequivalent vacua.


\subsubsection{Gaugino condensation and IASD flux}


To connect the gauge theory dynamics to our supergravity solution,\footnote{A precise matching between supergravity and gauge theory requires having a smooth solution, which we lack at present. Our discussion here will be qualitative, and limited to showing that the family of supersymmetric solutions found above has the required ingredients to encode gaugino condensation.} we use the results of~\cite{BDKKM}, which showed that gaugino condensation on D7-branes sources imaginary anti-self-dual (IASD) flux in the space
surrounding the branes.  Using the classical DBI coupling between D7-brane gauginos $\lambda$ and bulk fluxes~\cite{CIU},
\begin{equation} \label{CIUterm}
{\cal{L}} \ \supset \ \frac{a}{\alpha^{\prime 2}}\int\limits_{\mathbb P^2} \sqrt{g}\, G_{3} \cdot
\Omega \, \bar\lambda\bar\lambda \, + c.c.\,,
\end{equation} the flux equation of motion, expanded around a background containing exclusively imaginary self-dual fluxes, was found to be~\cite{BDKKM}
\begin{equation}\label{sourceterm}
{\rm{d}}\Bigl[e^{4A}\,(\star\, G_3-i\, G_3)\Bigr] =  \frac{4i a \kappa_{10}^2}{g_{s}\alpha^{\prime 2}}\,{\rm{d}}\Bigl[
\lambda\lambda\,\bar{\Omega}\,\delta(\mathbb P^2)\Bigr]  \,.
\end{equation}
Here $\star$ is the six-dimensional Hodge star, $\delta(\mathbb P^2)$ is a delta-function localizing on $\mathbb P^2$,
and $a$ is a dimensionless constant.

The nonzero expectation value (\ref{eq:vev-gaugino}) provides a localized source for flux in ten dimensions via (\ref{sourceterm}), and the resulting flux is IASD
with Hodge type $(1,2)$. Compelling evidence for this proposal comes from the fact that for
D7-branes wrapping a given four-cycle in a local geometry, the $G_{1,2}$ flux background induced by the coupling
(\ref{sourceterm}) precisely encodes the superpotential of the four-dimensional gauge theory: a D3-brane probing the flux background
sourced
by (\ref{sourceterm}) experiences the superpotential derived upon reduction to four dimensions~\cite{BDKKM}.

The analysis of~\cite{BDKKM} was performed in an expansion around type B backgrounds~\cite{Giddings:2001yu}, but focused on the theory of probe D3-branes\footnote{Consideration of more general D-brane probes~\cite{calibration1, calibration2} of our solutions is likely to lead to substantial physical insight, but is beyond the scope of the present work.}
and
consistently neglected perturbations to the metric and dilaton: such perturbations are clearly present as a consequence of the IASD flux sourced by the gaugino condensate, but do not contribute to the D3-brane scalar potential until third order.  However, it was natural to expect that the full solution of all the equations of
motion would unite the proposals of~\cite{Koerber:2007xk}, in which the background is a generalized complex geometry, and of \cite{BDKKM}, in which
$G_{1,2}$ flux plays the central role.  In this work we have exhibited a solution with dynamic $SU(2)$ structure that crucially involves $G_{1,2}$ flux, thereby illuminating the relationship between~\cite{Koerber:2007xk} and~\cite{BDKKM}.


\subsubsection{R-symmetry breaking and domain walls}


Having explained the relation between three-form fluxes and gaugino condensation, let us discuss how the pattern of R-symmetry breaking described above may be encoded in the supergravity solution. Referring to Table~\ref{table:Scharge}, we see that $R(\Theta)=+2$ and $R(\Omega_2)=0$. Comparing (\ref{eqn:Theta_c1}, \ref{eqn:Omega_ansatz}) with the geometric action of $U(1)_{\psi}$ (appropriately normalized as in \S\ref{subsec:CYgeom}), we see that $c_1$ and $e^{i \xi}$ must carry R-charges $+2$ and $-2$ respectively.
This suggests that, in a four-dimensional off-shell formulation (where the equations of motion of the effective action reproduce the ten-dimensional equations), some combination of $c_1$ and $e^{i \xi}$ becomes a fluctuating space-time field with nonvanishing R-charge, whose expectation value spontaneously breaks the exact  R-symmetry to $\mathbb Z_2$.

Gaugino condensation has a similar off-shell description in terms of the glueball superfield
$S = -\frac{1}{32 \pi^2} {\rm tr}\,W_\alpha W^\alpha$ and the Veneziano-Yankielowicz superpotential~\cite{Veneziano:1982ah}. On-shell this field acquires a nonzero expectation value proportional to (\ref{eq:vev-gaugino}) and reproduces the nonperturbative superpotential. It is natural to conjecture that the combination of $c_1$ and $e^{i \xi}$ mentioned above
is dual to $S$; then the on-shell superpotential (\ref{eqn:KMsuperpotential}, \ref{eqn:onshell_superpotential}) would have to agree with the gaugino-condensate superpotential.\footnote{For D5-branes or D6-branes in the conifold this matching was obtained in the large N duality of~\cite{Vafa:2000wi}.}


We expect the appearance of domain walls due to the spontaneous breaking of the exact R-symmetry $\mathbb{Z}_{2 \mathfrak{c_2}} \to \mathbb{Z}_2$;
these
should correspond to wrapped branes in the gravity solution. We now argue that D3-branes wrapping a loop inside the $S^5/\mathbb{Z}_3$ have the right properties to be domain walls in our solution.

The $S^5/\mathbb{Z}_3$ has fundamental group $\mathbb{Z}_3$ generated by a loop where $\psi$ runs from $0 \to 2 \pi$. However, due to the involution, a D3-brane wrapping from $0 \to \pi$ is permitted, where the two ends are identified under the involution $\psi \to \psi+\pi$. This corresponds to the generator of the $\mathbb{Z}_6$ fundamental group of the horizon in the downstairs geometry. We compute the tension of the domain wall by evaluating the DBI action for the brane in the probe approximation:
\be \label{eqn:walltension}
T_{\rm (wall)} = \mu_3 \oint \mathd s_{(6)}\, e^{2 A} =  \mu_3\, \pi\, r\, e^{2 (A-B)} = \mu_3\, \pi\, |c_1|\, \frac{|\Xi_++\Xi_-|}{2\, |\Xi_+\, \Xi_-|^{1/2}} \ge \mu_3\, \pi\, |c_1|\,,
\ee
where the bound is saturated if and only if $\Xi_+ = \Xi_-$ at the location of the domain wall. Thus, the tension is minimized at a $\varphi = \pi/2$ locus, in which case the wall is a half-BPS state~\cite{calibration2}. The analysis of \S\ref{subsec:validity} describes one example where the supergravity approximation is valid at such a locus.

The $\mathbb{Z}_6$ fundamental group in the downstairs geometry implies that the number of domain walls of this type is conserved modulo six. Thus, the gauge theory has six vacua, consistent with gaugino condensation in $SO(8)$ pure super Yang-Mills.  We anticipate that these vacua are related by exact R-symmetry transformations. The BPS domain wall tension (\ref{eqn:walltension}) can then be used to infer the precise value of the superpotential vev.
We postpone a more detailed study of these domain walls and other wrapped branes to a future work.



Finally, we recall that the extension of the gravity solution to the full $\mathbb P^2$ cone using the ansatz of \S\ref{sec:ansatz} reveals that the space-time becomes AdS, as explained in \S \ref{subsec:P2_susy}.
We have not obtained a satisfactory interpretation of this restriction from the viewpoint of the gauge theory.


\subsection{Applications}

Our solution has a range of interesting
applications, two of which we now describe.

\subsubsection{Ten-dimensional consistency of KKLT vacua}\label{KKLT}

One
application would be to study the ten-dimensional
consistency conditions for KKLT vacua.  As vacua of the
four-dimensional effective theory, these solutions are reasonably well
understood, but are known to violate constraints that emerge from the
ten-dimensional type IIB supergravity equations of motion with
classical sources.
Specifically, from the external Einstein equations and the five-form
Bianchi identity, one obtains
\begin{equation} \label{phieom}
\nabla^2 \Phi_- = \frac{1}{4}\, e^{8A} \left|\star\, G_3-i\, G_3\right|^2 + {\cal R}_4 +
e^{-4 A}\, | \nabla \Phi_-|^2  + {\cal S}_{\sf local}\, ,
\end{equation}
where ${\cal R}_4$ is the four-dimensional Ricci scalar, and \cite{Giddings:2001yu}
\begin{equation} \label{localsource}
{\cal S}_{\sf local} =
2\kappa_{10}^2\,e^{2A}\,\Biglp \frac{e^{2A}}{4}T^{m}_{m}-\frac{e^{-2A}}{4}T^{\mu}_{\mu}-\mu_3 \rho_3\Bigrp_{\sf
local}\,,
\end{equation}
where $T^{m}_n$ and $T^{\mu}_{\nu}$ are the internal and external components of the ten-dimensional stress-energy tensor $T^{M}_N$ (with indices raised by the unwarped metrics $g^{mn}$ and $h^{\mu\nu}$, respectively),
and $\rho_3$ is the D3-brane charge density.  An anti-D3-brane at
position $y_{0}$ provides a {\it{positive}} localized source,
\be
{\cal S}_{\sf local}^{\overline{D3}} = 4\, \kappa_{10}^2\, \mu_3\, e^{8 A}\,\delta^{6}(y-y_{0})\, ,
\ee
whereas D3-branes, O3-planes, and O7-planes, like all other local
sources allowed in the solutions of \cite{Giddings:2001yu}, provide a vanishing
contribution to ${\cal S}_{\sf local}$.
Then, noting that the integral of the left-hand side of (\ref{phieom})
over a compact space vanishes, one learns that a de Sitter solution is
possible only if a suitable localized negative contribution to the right-hand
side is present.  We will denote such a contribution as $\rho_{-}(y)$.

We emphasize that negative tension alone does not suffice to produce a
contribution to $\rho_{-}$, as is evident from the fact that O3-planes
and O7-planes do not contribute.
Localized classical sources that do contribute to $\rho_{-}$ include
anti-O3-planes and O5-planes \cite{Giddings:2001yu}, but to our knowledge such
objects have not played a role in the construction of consistent de
Sitter vacua in the framework of \cite{Kachru:2003aw}.

A natural guess is that the four-dimensional nonperturbative effects
that led to stability in the effective theory -- i.e., gaugino
condensation on D7-branes -- will provide new sources in the
ten-dimensional equations of motion.  In fact, the stress tensor
contribution given in (\ref{phieom}) arises from varying the classical
action for {\it{bosonic}} fields with respect to the metric.  The
stress tensor by definition involves the variation of the complete
action, but fermionic expectation values vanish in classical vacua,
and the corresponding contributions may then be omitted.  However, in
the solutions we have considered, the fermion bilinear
$\lambda\lambda$ plays an essential role, and contributions
proportional to $\langle\lambda\lambda\rangle$ should be retained.

Specifically, to incorporate the effects of gaugino condensation one
should also vary the coupling (\ref{CIUterm})
with respect to the metric.  The result is a new contribution to
(\ref{localsource})
that is
{\it{negative}} and proportional to
$|\langle\lambda\lambda\rangle|^2$.  We speculate that
this negative contribution could suffice to establish that KKLT
vacua can be lifted to consistent ten-dimensional solutions, but we
leave a thorough investigation of this point for the future.

\subsubsection{On del Pezzo transitions with seven-branes}

A more speculative application of our result is to a geometric
transition for seven-branes.  It would be interesting to understand when a divisor wrapped by seven-branes can be contracted in such a way that the resulting singularity can be deformed.  The only divisors in a Calabi-Yau threefold that admit birational contraction followed by deformation to a new Calabi-Yau are the
del Pezzo surfaces $\mathbb{P}^{1}\times \mathbb{P}^{1}$ and $dP_{k}$, $k\ge 2$ \cite{Rossi, Gross}.\footnote{Strictly speaking, del Pezzo surfaces are the only possible exceptional divisors for `primitive' contractions, from which more general contractions can be constructed \cite{Gross}.}  (The del Pezzo surfaces $dP_{0}=\mathbb{P}^2$ and $dP_{1}$ can be contracted, but the  resulting singular varieties cannot be deformed to
smooth Calabi-Yau threefolds.)

We would like to understand when del Pezzo transitions can occur for divisors wrapped by seven-branes, motivated by the rich physics of geometric transitions involving D5-branes.
The role of gaugino condensation in the conventional D5-brane geometric transition \cite{Vafa:2000wi}
is well understood, and strongly suggests that
for seven-branes it is also important to characterize the effect of gaugino condensation in the geometry.  Our result prepares the tools
for such an analysis, but exploring a seven-brane geometric transition in detail is beyond the scope of this work.  We also observe that in any case where a del Pezzo transition with seven-branes {\it{is}} possible, so that a smooth geometry is obtained after the deformation, the resulting absence of a local source for $\rho_{-}$ as described in \S\ref{KKLT} presents an obstacle to obtaining consistent de Sitter vacua.

\section{Conclusions}\label{sec:concl}

We have obtained a family of exact, noncompact, supersymmetric solutions of type IIB supergravity with dynamic $SU(2)$ structure.  The core of each solution is a stack of four D7-branes atop an O7-plane on a flat non-compact four-cycle.  We argued that  this solution describes the region near a small patch of a compact, rigid four-cycle, within which we have assumed rotational symmetry.
We gave strong evidence for this claim by presenting an ansatz for a corresponding configuration of four D7-branes and an O7-plane wrapping the $\mathbb{P}^2$ base of the simplest del Pezzo cone, and showing that in the near-stack limit our flat ansatz is recovered.
Decompactifying the four-cycle destroys information about induced charges and sends the seven-brane gauge coupling to zero, while drastically simplifying the equations of motion.  By comparing the flat-space solution to the $\mathbb{P}^2$ cone configuration, we were able to interpret certain aspects of our
solution as arising from seven-brane gauge dynamics or from induced D3-brane charge and tension.

For compact, rigid four-cycles, the seven-brane gauge theory undergoes gaugino condensation at low energies.  In this work we have identified
a class of exact solutions as candidates for the ten-dimensional backreaction of seven-brane gaugino condensation.  Our approach was
strictly ten-dimensional and did not incorporate nonperturbative source terms, in contrast to, but not in contradiction with, \cite{Koerber:2007xk},\footnote{See e.g.\ \cite{Polchinski:2000uf,Grana:2000jj,Lopes Cardoso:2004ni} for earlier connections between gaugino condensation and generalized complex geometry, and \cite{BDKKM} for evidence that gaugino couplings to flux source ten-dimensional deformations.} which proposed that gaugino condensation provides a localized source term that induces a deformation to a  generalized complex geometry.  It would be valuable to understand the relationship between these approaches.

Although the seven-brane charge necessarily vanishes in our solutions, the three-brane charge and tension induced on seven-branes wrapping $\mathbb{P}^2$
can be negative, so that at short distances one expects singular behavior typical of orientifolds.  We indeed find a singularity with divergent warp factor near the seven-branes. It would be very interesting to understand if this singularity is ultimately removed by strong gauge dynamics on the seven-branes.

Stacks of seven-branes wrapping rigid four-cycles are ubiquitous in type IIB compactifications, and understanding their effects in ten-dimensional supergravity is an important step toward characterizing the resulting four-dimensional effective theories.  In particular, decoupling arguments analogous to \cite{RS} that invoke extradimensional locality require a ten-dimensional description, and the effective theory of D3-branes is most efficiently described
by geometrizing seven-brane gauge dynamics \cite{BDKKM}, as we have done here.  The configuration we have presented is arguably the simplest nontrivial example of the backreaction of seven-brane nonperturbative effects, because the seven-brane charges vanish and the four-cycle is highly symmetric.  Our approach can be extended to configurations with less symmetry, such as seven-branes wrapping the base of a suitable del Pezzo cone; we argued that analogous solutions exist for the $\mathbb{P}^2$ cone.  It would be very interesting to construct additional examples and explore their implications, both as windows into the dynamics of the seven-brane gauge theory, and as descriptions of local regions of stabilized type IIB compactifications.

\subsection*{Acknowledgements}
We are grateful to D. Baumann, M. Berg, P. C\'amara, A. Collinucci, F. Denef, A. Dymarsky, S. Franco, T. Grimm, S. Kachru, L. Martucci, G. Moore, E. Silverstein, T. Weigand, and T. Wrase for helpful discussions.  We thank D. Baumann, S. Kachru,
J.~Liu, L. Martucci, and P.~Szepietowski for useful comments on the manuscript. L.M. specially thanks T. Weigand for valuable discussions and correspondence.  G.T. is particularly grateful to S. Kachru for many enlightening conversations. The research of L.M. was supported by the Alfred P. Sloan Foundation and by the NSF under grant PHY-0757868. The research of B.H. was supported in part by a Cornell University Olin Fellowship.  B.H. and L.M. gratefully acknowledge support for this work by the Swedish Foundation for International Cooperation in Research and Higher Education. G.T. is supported by the US DOE under contract number DE-AC02-76SF00515 at SLAC.  We thank the organizers of String Phenomenology 2010 for providing a stimulating environment for a portion of this work.  G.T. would like to thank the KITP, where part of this work was done, for hospitality.

Portions of this document were prepared with the help of \TeXmacs, free software available at \url{http://www.texmacs.org}.

\appendix

\section{Equations of Motion for the Near-stack Ansatz} \label{appendix:eoms}

We now present the general equations of motion for the warped ansatz (\ref{eqn:10dmetric}, \ref{eqn:F5ansatz}), and then specialize to the near-stack limit.
Rewriting the covariant action (\ref{eqn:covarSUGRA}) in terms of a four-dimensional Lagrangian density,
\be
  S = \int \mathd^4\, x\, \sqrt{- h}\, \mathcal{L} \,,
\ee
where $\mathcal{L}$ is given by an integral over the internal space, and applying
the warped ansatz (\ref{eqn:10dmetric}, \ref{eqn:F5ansatz}),{\footnote{As usual, there are some surmountable subtleties relating to the
self-duality of $\tilde{F}_5$.}} we
find the vacuum Lagrangian:
\begin{eqnarray}\label{eqn:vacLagrangian}
  \mathcal{L} & = & \frac{1}{2 \kappa_{10}^2}  \int \mathd^6 y \sqrt{g}
  \left[ {\cal{R}} + \frac{1}{2} e^{- 8 A} (\nabla \alpha)^2 - 8 (\nabla A)^2 + 4
  \Lambda e^{- 4 A} - \frac{1}{2}  \left(\frac{1}{\tau_2^2} |\mathd \tau|^2  + e^{4 A} |G_3 |^2 \right)
  \right] \nonumber\\
  &  & - \frac{i}{4 \kappa_{10}^2}  \int \alpha\, G_3 \wedge G_3^{\star}\,, \label{eqn:vaclagrangian}
\end{eqnarray}
where $\Lambda={\cal{R}}_{(4)}/4$ is the four-dimensional cosmological constant, $\tau_2 = \mathrm{Im}\, \tau$,
contractions are made using the unwarped metric $g_{m n}$, and $G_3
= \mathD_-\, \mathcal{A}_2$, so that $\mathD_-\, G_3 = 0$.
The corresponding equations of motion are:
\begin{eqnarray}
  \mathd \left( e^{- 8 A} \star_6 \mathd \alpha \right)  =  - \frac{i}{2}\,G_3 \wedge G_3^{\star} & \;\;,\;\; &
  \nabla^2 A = \frac{1}{8} e^{4 A} |G_3 |^2 + \frac{1}{4} e^{- 8 A}
  (\nabla \alpha)^2 + \Lambda e^{- 4 A}\,, \\
  \mathD_-  \left( e^{4 A} \star_6 G_3^{\star} \right) =  - i\, \mathd \alpha
  \wedge G_3^{\star} & \;\;,\;\; &
  \mathD \star_6 \left(\frac{1}{\tau_2}\, \mathd \bar{\tau} \right)  =  \frac{i}{2} e^{4 A} G_3^{\star} \wedge \star_6
  G_3^{\star}\,,
\end{eqnarray}
\be
R_{m n} = 8\, \nabla_m A\, \nabla_n A - \frac{1}{2} e^{- 8 A}\, \nabla_m
  \alpha\, \nabla_n \alpha
  +\frac{1}{4 \tau_2^2}  \left[ \nabla_m \tau
  \nabla_n  \bar{\tau} + c.c. \right]+\frac{1}{2} e^{4 A}\,  \hat{T}_{m n} - \Lambda e^{- 4 A} g_{m n}\,,
\ee
where
\be
  \hat{T}^m_n = \frac{1}{4} (G^{m p q}  \bar{G}_{n p q} + \bar{G}^{m p q}
  G_{n p q}) - \frac{1}{12}  \bar{G}^{p q r} G_{p q r} \delta^m_n\,.
\ee
To work out the Einstein equations, we compute $\hat{T}$ in a complex basis. It is straightforward to check that
$\hat{T}^{\mu}_{\nu} = 0$ if $\mu$ and $\nu$ are both holomorphic indices; this is
a consequence of the primitivity of $G_3$. Of the mixed,
$\hat{T}^{\bar{\mu}}_{\nu}$ components, all except $\hat{T}^{\bar{z}}_z= (\hat{T}^z_{\bar{z}})^{\star}$ must
vanish by symmetry, and we find
\be
 \hat{T}_{z z} = 4\, e^{- 4 C}  \frac{\bar{z}}{z}  \left[ g^{3, 0} \bar{g}^{2, 1} + g^{1, 2}  \bar{g}^{0, 3} \right]\,.
\ee
After a straightforward computation, we find the Ricci components
\bea
R_{z z} = - \frac{\bar{z}}{z}  \left[ C'' + 4 B' C' + (C')^2 - \frac{1}{r} C' \right] & \;\;,\;\; &
R_{z \bar{z}} = \left[ B'' - C'' + \frac{1}{r} (B' - C') - (C')^2 \right]\,, \nonumber \\
R_{u^i \bar{u}^{\bar{j}}}= - \frac{1}{2} e^{4 B + 2 C}  \left[ C'' + \frac{1}{r} C' + 4 (C')^2 \right] \delta_{i \bar{j}} \,. & \;\;\,\;\; &
\eea
Using these formulae, one can write down the Einstein equations in terms of $B$ and $C$. Applying the remaining equations of motion to the ansatz of \S\ref{subsec:flat} in like fashion, we find four real second order equations of motion for $A$, $\alpha$, $B$, and $C$, along with one complex second order equation of motion for $\tau$, four real first order equations of motion for the $g_{p,q}$, and one complex constraint coming from the $r, \psi$ component of the Einstein equations.

The $\alpha$ and $A$ equations of motion are
\be \label{eqn:alphaeom}
  \frac{1}{r}  \frac{\mathd}{\mathd r}  \left( r e^{4 (C - 2 A)} \alpha'
  \right)  =  4 \sum_{p = 0}^3 (- 1)^p |g^{p, 3 - p} |^2 \,,
\ee
\be \label{eqn:Aeom}
  \frac{1}{r}  \frac{\mathd}{\mathd r}  \left( r e^{4 C} A' \right)  =  e^{4
  A}  \sum_{p = 0}^3 |g^{p, 3 - p} |^2 + \frac{1}{4} e^{4 (C - 2 A)}
  (\alpha')^2 + \Lambda e^{4 (C - B - A)} \,,
\ee
where primes denote derivatives with respect to $r$. The $\tau$ equation of motion is
\be \label{eqn:taueom}
  \frac{1}{r \tau_2}  \frac{\mathd}{\mathd r}  \left( r e^{4 C} \tau' \right)
  + \frac{i}{\tau_2^2} e^{4 C} (\tau')^2 = - 8 i e^{4 A}  \left( g^{3, 0}
  g^{0, 3} + g^{2, 1} g^{1, 2} \right) \,.
\ee
The $G_3$ equations of motion and Bianchi identities are
\be \label{eqn:g30eom}
  \frac{1}{r \sqrt{\tau_2}}  \frac{\mathd}{\mathd r}  \left[ r \sqrt{\tau_2}
  e^{4 A} g^{3, 0} \right] + \frac{i \tau'}{2 \tau_2} e^{4 A}  \left[ g^{3, 0}
  + \bar{g}^{1, 2} \right] = \frac{e^{4 A}}{r}  g^{3, 0} + \frac{1}{2}
  (e^{4 A} - \alpha)'  \left[ g^{3, 0} - g^{2, 1} \right],
\ee
\be \label{eqn:g21eom}
  \frac{1}{r \sqrt{\tau_2}}  \frac{\mathd}{\mathd r}  \left[ r \sqrt{\tau_2}
  e^{4 A} g^{2, 1} \right] + \frac{i \tau'}{2 \tau_2} e^{4 A}  \left[ g^{2, 1}
  + \bar{g}^{0, 3} \right] = - \frac{e^{4 A}}{r}  g^{2, 1} - \frac{1}{2}
  (e^{4 A} + \alpha)'  \left[ g^{3, 0} - g^{2, 1} \right],
\ee
\be \label{eqn:g12eom}
  \frac{1}{r \sqrt{\tau_2}}  \frac{\mathd}{\mathd r}  \left[ r \sqrt{\tau_2}
  e^{4 A} g^{1, 2} \right] + \frac{i \tau'}{2 \tau_2} e^{4 A}  \left[
  \bar{g}^{3, 0} + g^{1, 2} \right] = - \frac{e^{4 A}}{r}  g^{1, 2} -
  \frac{1}{2} (e^{4 A} - \alpha)'  \left[ g^{0, 3} - g^{1, 2} \right],
\ee
\be \label{eqn:g03eom}
  \frac{1}{r \sqrt{\tau_2}}  \frac{\mathd}{\mathd r}  \left[ r \sqrt{\tau_2}
  e^{4 A} g^{0, 3} \right] + \frac{i \tau'}{2 \tau_2} e^{4 A}  \left[
  \bar{g}^{2, 1} + g^{0, 3} \right] = \frac{e^{4 A}}{r}  g^{0, 3} +
  \frac{1}{2} (e^{4 A} + \alpha)'  \left[ g^{0, 3} - g^{1, 2} \right].
\ee
The $B$ and $C$ equations of motion are
\be \label{eqn:Ceom}
  C'' + \frac{1}{r} C' + 4 (C')^2  =  \Lambda e^{- 4 (A + B)}\,,
\ee
\be \label{eqn:Beom}
  B'' + \frac{1}{r} B' + 3 (C')^2 = 2 (A')^2 - \frac{1}{8} e^{- 8 A}
  (\alpha')^2 + \frac{1}{8 \tau_2^2} | \tau' |^2 + \frac{1}{2} e^{- 4 (A + B)}
  \Lambda \,,
\ee
and the constraint takes the form
\begin{eqnarray} \label{eqn:BCcons}
  C'  \left[ \frac{2}{r} + 3 C' - 4 B' \right] & = & 2 (A')^2 - \frac{1}{8}
  e^{- 8 A} (\alpha')^2 + \frac{1}{8 \tau_2^2} | \tau' |^2 \nonumber \\
  & & + 2 e^{4 (A - C)}
  \left( g_{3, 0}\,  \bar{g}_{2, 1} + g_{1, 2}\,  \bar{g}_{0, 3} \right) + e^{- 4 (A + B)} \Lambda \,.
\end{eqnarray}

\bibliographystyle{JHEP}
\renewcommand{\refname}{Bibliography}
\addcontentsline{toc}{section}{References}
\providecommand{\href}[2]{#2}\begingroup\raggedright

\end{document}